\documentclass[twocolumn]{article}

\usepackage[english]{babel}

\usepackage[letterpaper,top=2cm,bottom=2cm,left=2cm,right=2cm,marginparwidth=1.75cm]{geometry}

\usepackage{footmisc}
\usepackage{appendix}
\usepackage{amssymb}
\usepackage{amsmath}
\usepackage{multirow}
\usepackage{appendix}
\usepackage{float}
\usepackage{graphbox}
\usepackage{authblk}
\usepackage[colorlinks=true, allcolors=blue]{hyperref}
\usepackage{subcaption}
\usepackage{indentfirst}
\usepackage{graphicx}
\usepackage{comment}
\usepackage{multirow}
\usepackage{xcolor}
\usepackage[figurename=Fig.]{caption}
\usepackage{cite}
\usepackage{amsfonts}
\usepackage{algorithm}
\usepackage{algorithmic}
\usepackage{rotating} 
\usepackage{booktabs}

\title{Improving Generalization and Trainability of Quantum Eigensolvers via Graph Neural Encoding}
\author[1]{\normalsize Jungyun Lee}
\author[1,2,3,*]{\normalsize Daniel K. Park}
\affil[1]{\small \textit{Department of Statistics and Data Science, Yonsei University, Seoul, Republic of Korea}}
\affil[2]{\small \textit{Department of Applied Statistics, Yonsei University, Seoul, Republic of Korea}}
\affil[3]{\small \textit{Department of Quantum Information, Yonsei University, Seoul, Republic of Korea}}

\date{}

\begin{document}

\maketitle
{\let\thefootnote\relax\footnotetext{\small * : corresponding author}}
{\let\thefootnote\relax\footnotetext{\small Email addresses : dkd.park@yonsei.ac.kr}}

\begin{abstract}

Determining the ground state of a many-body Hamiltonian is a central problem across physics, chemistry, and combinatorial optimization, yet it is often classically intractable due to the exponential growth of Hilbert space with system size. Even on fault-tolerant quantum computers, quantum algorithms with convergence guarantees---such as quantum phase estimation and quantum subspace methods---require an initial state with sufficiently large overlap with the true ground state to be effective. Variational quantum eigensolvers (VQEs) are natural candidates for preparing such states; however, standard VQEs typically exhibit poor generalization, requiring retraining for each Hamiltonian instance, and often suffer from barren plateaus, where gradients can vanish exponentially with circuit depth and system size. To address these limitations, we propose an end-to-end representation learning framework that combines a graph autoencoder with a classical neural network to generate VQE parameters that generalize across Hamiltonian instances. By encoding interaction topology and coupling structure, the proposed model produces high-overlap initial states without instance-specific optimization. Through extensive numerical experiments on families of one- and two-local Hamiltonians, we demonstrate improved generalization and trainability, manifested as reduced test error and a significantly milder decay of gradient variance. We further show that our method substantially accelerates convergence in quantum subspace-based eigensolvers, highlighting its practical impact for downstream quantum algorithms.
\end{abstract}

\section{Introduction}
Determining the ground state of a Hamiltonian and its corresponding energy is of fundamental importance in a wide range of problems across physics, chemistry, and combinatorial optimization~\cite{feynman1982simulating, levine2009quantum,farhi2014quantumapproximateoptimizationalgorithm,lucas2014ising,lee2023evaluating}. 
Identifying the ground state of the Hamiltonian in general is known to be classically intractable, as the computational complexity of performing exact diagonalization of the Hamiltonian grows exponentially with system size~\cite{lloyd1996universal, bittel2021training}. As a means to mitigate this exponential growth in computational resources, variational methods based on the Rayleigh–Ritz principle restrict the search to a subspace of the full Hilbert space. However, many studies demonstrate that the classical variational approaches still struggle to represent and optimize highly entangled many-body states~\cite{harrow2017quantum, boixo2018characterizing,yuan2019theory}. 

These classical limitations have motivated the exploration of quantum computing as a potential alternative~\cite{feynman1982simulating}. It has been shown that quantum computers can efficiently simulate local quantum systems using, for example, the Trotter–Suzuki decomposition~\cite{lloyd1996universal,trotter1959product,suzuki1993improved}. Building on this foundation, early quantum algorithms, such as the quantum phase estimation (QPE), were developed to access ground-state information by probabilistically projecting an initial state onto Hamiltonian eigenstates~\cite{kitaev1995quantummeasurementsabelianstabilizer, abrams1999quantum, aspuru2005simulated}. While QPE offers exponential advantages over classical counterparts in principle, it requires deep quantum circuits with high-fidelity quantum gates, preventing its scalable deployment on current noisy intermediate-scale quantum (NISQ) devices~\cite{preskill2018quantum, stair2021simulating}. Moreover, even on fault-tolerant quantum computers, the effectiveness of QPE fundamentally relies on the overlap between the initial state and the true ground state of the Hamiltonian, $\Omega = |\langle \psi_{\text{initial}}|\psi_{\text{ground}}\rangle|^2$, since number of measurement repetitions required by QPE scales as $\mathcal{O}(1/\Omega)$. For a generic initial state, $\Omega$ typically decreases exponentially with the number of qubits $n$, causing the expected number of repetitions to grow exponentially and thereby undermining the anticipated quantum advantage~\cite{shang2024polynomialtimedissipationbasedquantumalgorithm, dong2022ground}. Consequently, achieving initial states for which $\Omega$ decays at most polynomially with system size remains a central challenge in practice.

To alleviate these challenges, quantum subspace methods (QSMs), such as Krylov quantum diagonalization (KQD)~\cite{KQD_1, KQD_2} and  sample-based Krylov quantum diagonalization (SKQD)~\cite{SKQD}, have emerged as an alternative approach for estimating spectral properties on near-term and early fault-tolerant quantum devices.
These methods provide rigorous convergence guarantees for ground-state energy estimation.
However, these guarantees typically assume an initial state with at least inverse-polynomial overlap with the true ground state as in QPE, and sample-based variants additionally require sparsity of the ground state in the measurement basis~\cite{SQD, SKQD}.


Another prominent algorithm for NISQ devices is the variational quantum eigensolver (VQE)~\cite{peruzzo2014variational}, which is based on the Rayleigh--Ritz variational principle. VQEs employ parameterized quantum circuits (ans\"{a}tze) whose parameters are optimized using classical algorithms to approximate the ground state and its corresponding energy~\cite{mcclean2016theory, cerezo2021variational}. This hybrid quantum--classical framework has been applied across optimization, quantum simulation, and quantum machine learning~\cite{farhi2014quantumapproximateoptimizationalgorithm, perezramirez2024variationalquantumalgorithmscombinatorial, yuan2019theory, tilly2022variational, watad2024variational, wittek2014quantum, benedetti2019parameterized, hur2022quantum}. Although VQEs are heuristic and lack rigorous convergence guarantees, they can serve as effective state-preparation routines by producing initial states with nontrivial overlap for downstream algorithms such as QPE and QSMs~\cite{QSCI}, thereby reducing sampling overhead. As a result, VQEs remain relevant both in the NISQ era and in future fault-tolerant regimes~\cite{zimborás2025mythsquantumcomputationfault}.

Despite early optimism, the practical deployment of VQEs faces significant challenges. One major obstacle is the barren plateaus (BP) phenomenon~\cite{mcclean2018barren,larocca2025barren}, where gradients of the cost function vanish exponentially with system size or circuit depth, severely impeding optimization. In addition, standard VQEs lack generalization capability: they must be retrained from scratch for each Hamiltonian instance, even when different instances share substantial structural similarities. This fundamental limitation incurs significant computational overhead for real-world applications.

To address these limitations, we develop an end-to-end framework that generalizes across diverse families of one- and two-local Hamiltonians by integrating a Hamiltonian-graph autoencoder with a classical neural network (NN) into VQE training. Once trained, the model directly generates initial states having sufficient overlap with the true ground state of previously unseen Hamiltonians, without instance-specific optimization. Our approach leverages graph neural encoding for Hamiltonian representation learning to accommodate a broad class of interaction structures, achieving strong generalization and improved robustness to BP effects compared to prior learning-based VQE methods, such as neural-network-encoded VQE (NNVQE)~\cite{NN-VQA}.

The central contributions of this work are as follows:
\begin{itemize}
    \item We introduce a general mapping protocol that encodes arbitrary one- and two-local Pauli Hamiltonians as scalable Hamiltonian-graphs with multi-dimensional node and edge features.
    
    \item We propose an edge-featured graph attention autoencoder (EGATE) that jointly learns node and edge embeddings in a fully unsupervised manner, enabling expressive and scalable representation learning for quantum systems.

    \item We develop a representation-learning-based VQE framework, EGATE-NNVQE, which leverages latent Hamiltonian embeddings to directly generate high-overlap initial states for previously unseen Hamiltonians without instance-specific optimization. The proposed approach exhibits significantly improved generalization and a milder decay of gradient variance compared to standard VQE and other learning-based VQE baselines, indicating enhanced robustness to BP effects.    

    \item We demonstrate that EGATE-NNVQE substantially improves the performance of SKQD, enabling reliable convergence with smaller Krylov subspaces and fewer measurement shots under realistic noise models.
\end{itemize}

\section{Related Works}
\label{sec:NNVQE}

Neural-network-encoded VQE (NNVQE), proposed in Ref.~\cite{NN-VQA} as a state-of-the-art learning-based VQE method, aims to reduce the instance-wise optimization overhead of standard VQE by learning an NN that maps Hamiltonian parameters to ansatz circuit parameters for a predefined family of Hamiltonians with a fixed operator structure.

As an illustrative example, Ref.~\cite{NN-VQA} applies NNVQE to the one-dimensional (1D) antiferromagnetic XXZ Heisenberg spin chain with an external magnetic field subject to periodic boundary conditions, 
\begin{equation}
    H(\lambda) = \sum_{i}\left(\sigma^x_i\sigma^x_{i+1} + \sigma^y_i\sigma^y_{i+1} +\lambda \sigma^z_i\sigma^z_{i+1}\right) + \Delta\sum_i \sigma^z_i,
    \label{eq:XXZ spin}
\end{equation}
where $\sigma^\alpha_i$ are the Pauli operator acting on site $i$, $\alpha \in \lbrace x,y,z\rbrace$, and $\lambda$ is the anisotropy parameter treated as tunable variable, and $\Delta$ is a fixed longitudinal field strength.

The NNVQE training procedure is summarized as follows :
\begin{enumerate}
    \item A classical NN takes a Hamiltonian tunable parameter $\lambda_j$ as input and outputs the ansatz parameter,  $\vec{\theta}_j = f_{\phi}(\lambda_j)$, where $\phi$ denotes network weights.
    \item The ansatz $U(\vec{\theta}_j)$ parameterized by $\vec{\theta_j}$ prepares an output state, $|\psi_j\rangle = U(\vec{\theta}_j)|0\rangle^{\otimes n} = U\left(f_{\phi}(\lambda_j)\right)|0\rangle^{\otimes n}$, where $|0\rangle^{\otimes n}$ is the initial state for $n$-qubits system. 
    \item For the training set of tunable parameters $\lbrace \lambda_j\rbrace$ the cost function $C(\phi)$ is the expectation of $H(\lambda_j)$,
    \begin{align}
    C(\phi) &= \sum_j \langle H(\lambda_j)\rangle \notag\\
    &= \sum_j\langle 0|U^{\dagger}\left(f_{\phi}(\lambda_j)\right) 
        H(\lambda_j) U\left(f_{\phi}(\lambda_j)\right)|0\rangle,
    \label{eq:NN-VQE cost}
\end{align}
which is minimized by a gradient-based optimizer to obtain optimal network weights $\phi^*$; The gradient can be calculated by the parameter-shift rule~\cite{mitarai2018quantum,schuld2019evaluating} on the quantum computer or by the NN (back-propagation via the ansatz parameters)~\cite{NN-VQA}.
\end{enumerate}

After training, the NNVQE model can infer ansatz parameters for previously unseen values $\lambda_\mathrm{new}$ via $\vec{\theta}_\mathrm{new} = f_{\phi^*}(\lambda_\mathrm{new})$, yielding an approximation of the ground state and its energy without additional optimization.
The approach can be extended to multi-parameter Hamiltonians by increasing the dimensionality of the NN input; for example, treating both $\lambda$ and $\Delta$ as tunable variables leads to a two-dimensional input $(\lambda_j,\Delta_j)$, while the overall training procedure remains unchanged.

Despite its conceptual appeal, NNVQE exhibits notable limitations in practice. In particular, its generalization performance deteriorates when test Hamiltonians fall outside the training range or when the system size increases. This behavior arises because NNVQE relies solely on low-dimensional tunable parameters as inputs, without access to the interaction topology or operator-level structure of the Hamiltonian. As a result, the model lacks sufficient inductive bias to support robust generalization across structurally diverse Hamiltonians. Moreover, the NNVQE does not provide additional mitigation of BP phenomena beyond what is observed in standard VQE. Taken together, these limitations substantially constrain the practical applicability of NNVQE to narrowly defined Hamiltonian families.

In the following section, we introduce a graph-based Hamiltonian representation learning framework that learns more informative, structure-aware embeddings in an unsupervised manner, addressing key limitations of NNVQE and enabling improved generalization. A detailed comparison of our approach with standard VQE and NNVQE is presented in Sec.~\ref{sec:result}.


\section{Model Description}
\label{sec:model description}
\subsection{Hamiltonian-Graph}
\begin{figure}[t]
\centering
\begin{tabular}{c c}
    \includegraphics[width=0.4\linewidth]{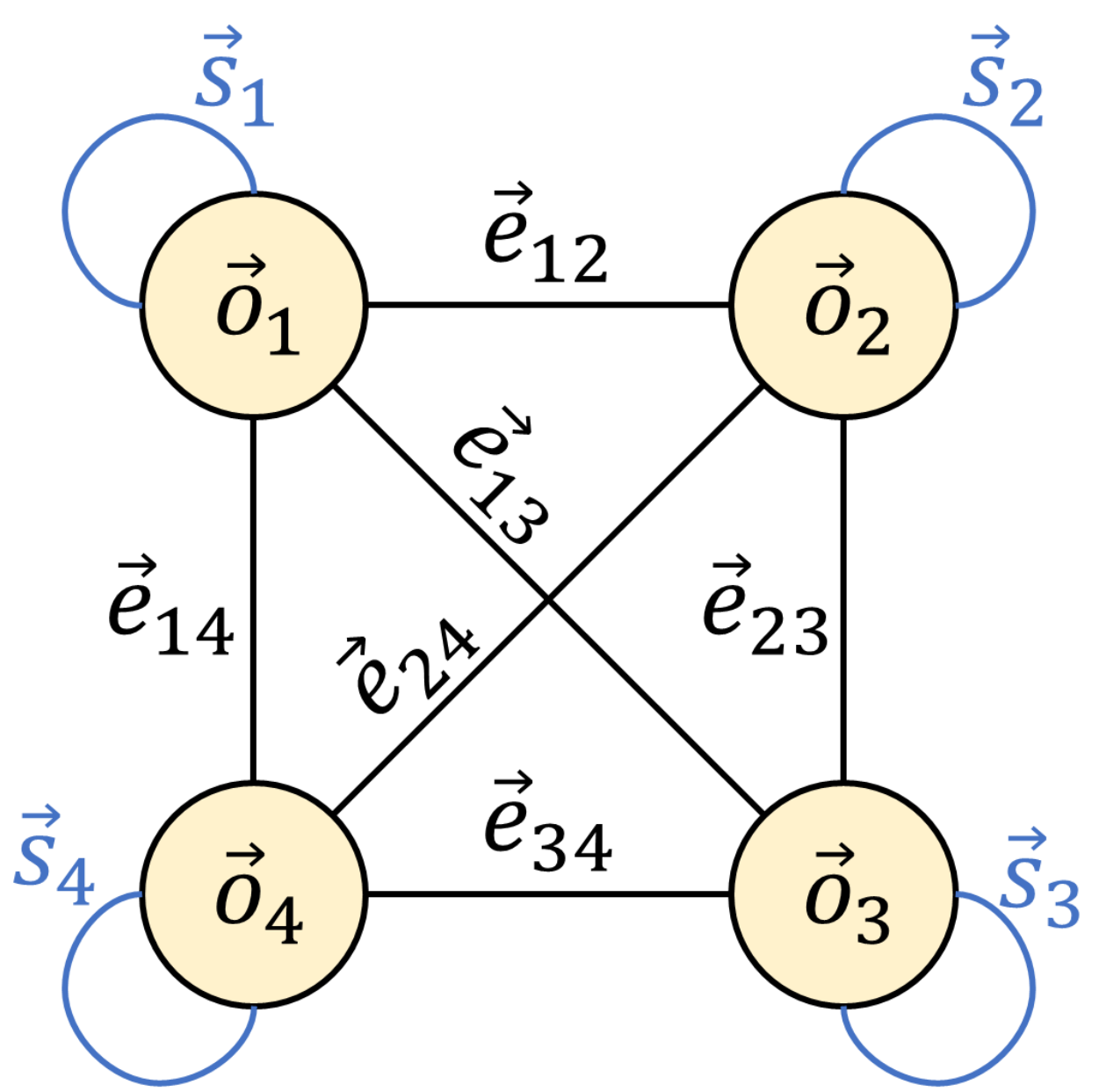} &
    \includegraphics[width=0.4\linewidth]{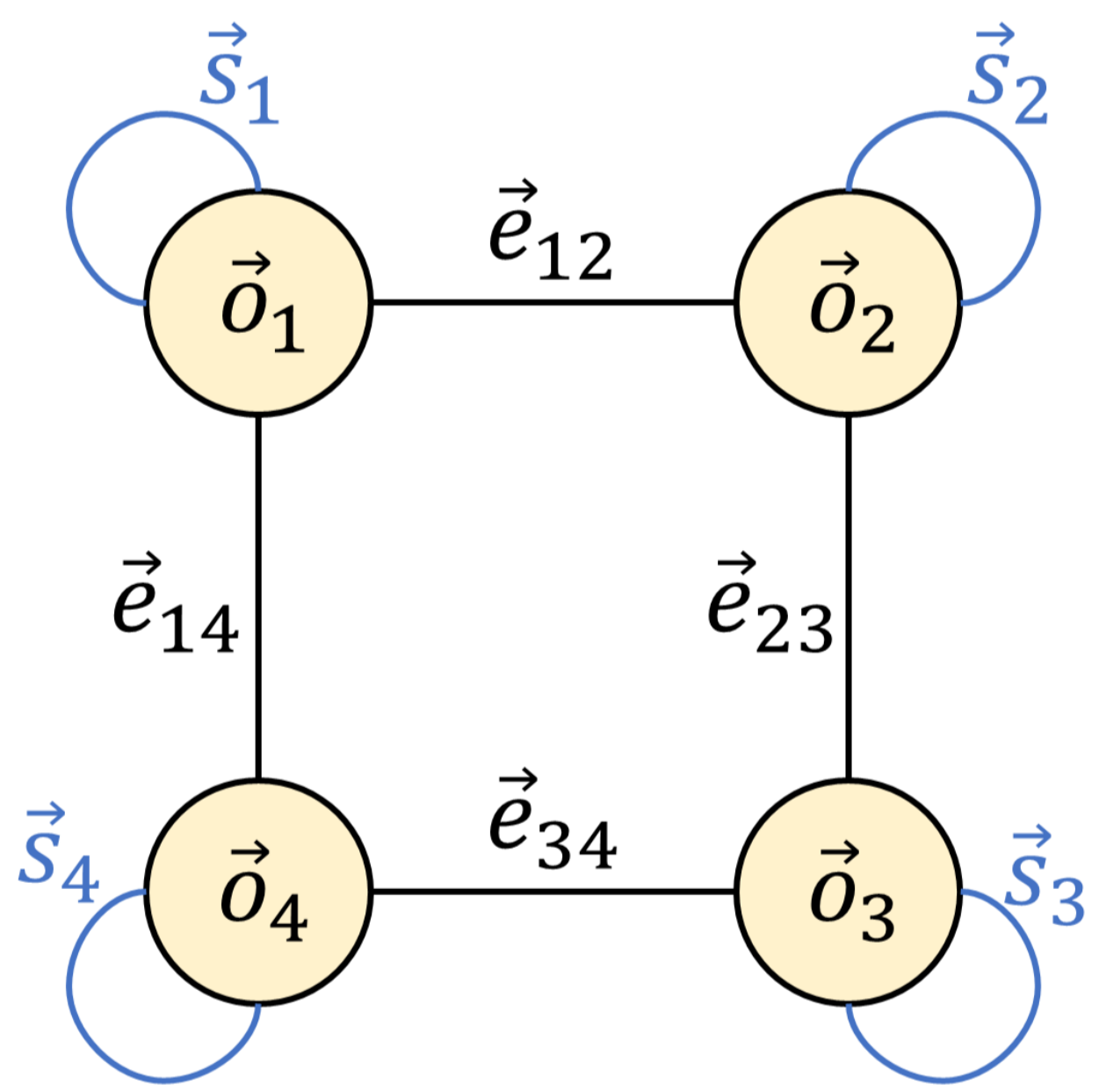} \\
    (a) & (b)    
  \end{tabular}
\caption{(a) Hamiltonian-graph for a general two-local 4-qubit Hamiltonian, as defined in Eq.~(\ref{eq:general two-local H}), and (b) Hamiltonian-graph for a 4-qubit 1D chain. If the two-local edge features in (b) are set to $\vec{e}_{ij} = (1, 1, \lambda)$ and the one-local edge features to $\vec{s}_i = (\Delta)$, the graph represents the 1D XXZ Heisenberg spin chain described in Eq.~(\ref{eq:XXZ spin})}
\label{fig:H-graph}
\end{figure}

In this section, we construct an undirected Hamiltonian-graph (H-graph) representation for a general one- and two-local Pauli Hamiltonian, which can be expressed as
\begin{equation}
H = \sum_{\langle i,j \rangle} \sum_{\alpha,\beta \in \lbrace x,y,z\rbrace} J_{ij}^{\alpha\beta} \sigma^\alpha_i \sigma^\beta_j
+\sum_i \sum_{\alpha \in \lbrace x,y,z\rbrace} K_i^\alpha \, \sigma^\alpha_i,
\label{eq:general two-local H}
\end{equation}
where $\sigma^\alpha_i$ and $\sigma^\beta_j$ denote Pauli operators acting on qubit $i$ and $j$, respectively, with $\alpha,\beta \in \lbrace x,y,z\rbrace$. 
Here $\langle i,j\rangle$ denotes the set of edges with $i<j$.
The coefficients $J_{ij}^{\alpha\beta}$ represent the two-local interaction strengths between qubits $i$ and $j$ for the $\sigma^\alpha_i \sigma^\beta_j$ terms, while $K_i^\alpha$ denotes the strength of the one-local interaction (e.g. due to an external field) acting on qubit $i$ along the $\alpha\in\lbrace x,y,z\rbrace$ direction. 
This formulation can represent any spin-1/2 Hamiltonian with at most two-local terms. Throughout the paper, we refer to Hamiltonians of the form in Eq.~(\ref{eq:general two-local H}) as general two-local Hamiltonians.

This general two-local Hamiltonian can be easily represented as a graph by identifying nodes with qubits and edges with pairwise couplings. To do so, we first assign each qubit $i$ a node feature vector $\vec{o}_i\in\mathbb{R}^{d_o}$ that encodes its site/structural information. There are several ways to represent the position or structure of the Hamiltonian node, for instance, using one-hot encoding, real-value, or coordinate value if the Hamiltonian is defined on a lattice. 
Additionally, one can utilize positional encodings~\cite{maskey2022generalizedlaplacianpositionalencoding, dwivedi2022graphneuralnetworkslearnable}, which inject structural information without explicit geometric coordinates such as lattice positions by assigning distinct embeddings to nodes purely based on their position or connectivity.
On the other hand, edge feature vector $\vec{e}_{ij}\in\mathbb{R}^{d_e}$ encodes the two-local couplings between nodes (qubits) $i$ and $j$, e.g., $\vec{e}_{ij} = \left(J_{ij}^{xx}, J_{ij}^{xy},\cdots, J_{ij}^{zz} \right)$. Thus, $d_e\le 9$.
To incorporate the one-local terms in Eq.~(\ref{eq:general two-local H}), we can introduce a self-loop edge for each node $i$. Each self-loop feature vector $\vec{s}_i=\left( K_i^x, K_i^y, K_i^z\right)\in\mathbb{R}^{d_s}$ with $d_s\leq 3$ encodes the local field strength acting on qubit $i$ along the $x$, $y$, and $z$ directions. 
Therefore, it can be absorbed into the node feature vector $\vec{o}_i$ of the corresponding site. Then, the dimensionality of the node feature vector changes to $\vec{o}_i\in\mathbb{R}^{d_o+d_s}$. This treatment encodes the effect of external fields while preserving the underlying physical interaction structure of the target Hamiltonian.

To utilize the H-graph for downstream tasks, we define a matrix representation of the H-graph $\textbf{G} = (\textbf{O}, \textbf{E})$ with $n$ nodes and $m$ edges. Here, the total node and edge matrix can be constructed by stacking each feature vector as a row; $\textbf{O} = [\vec{o}_i]_{i=1}^n\in\mathbb{R}^{n\times (d_o+d_s)}$ and $\textbf{E} = [\vec{e}_{ij}]_{\langle i,j \rangle}\in\mathbb{R}^{m\times d_e}$.
Importantly, since the H-graph places nodes in one-to-one correspondence with qubits, the number of nodes is the same as the number of qubits, $n$. Consequently, the representation requires quadratic memory in $n$ since $m=\binom{n}{2}=\mathcal{O}(n^2)$ in the worst case.

\begin{figure*}[t]
\centering
\includegraphics[width=0.8\textwidth]{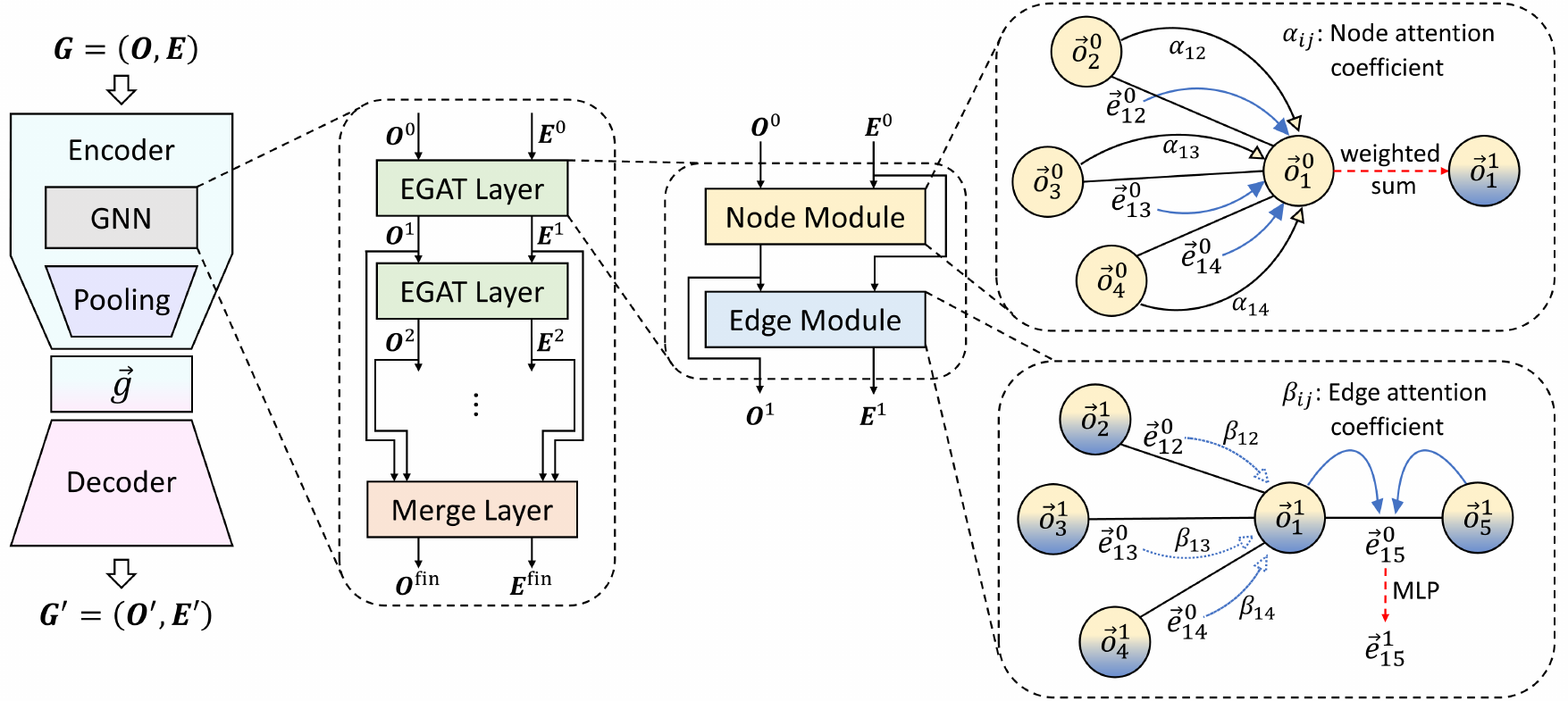} 
\caption{Architecture of EGATE and its components. The left panel illustrates the encoder-decoder structure of EGATE. The encoder consists of a GNN block for message passing followed by a pooling block that compresses the input H-graph $\textbf{G} = (\textbf{O}, \textbf{E})$ into a latent vector $\vec{g}$, while the decoder reconstructs the H-graph $\textbf{G}' = (\textbf{O}', \textbf{E}')$ from this latent representation. The center of the figure depicts the problem-specific GNN block, composed of stacked EGAT layers and a merge layer that jointly update node and edge features. The right panel details a single EGAT layer: the top and bottom schematics illustrate the update procedures of the node and edge modules, respectively. Here, $\alpha$ and $\beta$ denote attention coefficients. The corresponding pseudocode is provided in Algs.~\ref{appendix:alg:node module} and~\ref{appendix:alg:edge module}.}
\label{fig:model}
\end{figure*}

\subsection{Edge-Featured Graph Attention Autoencoder}
\label{sec:EGATE}
This section presents an edge-featured graph attention autoencoder (EGATE), a graph autoencoder (GAE) specifically designed to learn expressive and compact representations of H-graphs. 
During training, EGATE learns low-dimensional latent representations of H-graphs that compresses their structural and physical information by minimizing a reconstruction loss.
Once trained, we discard the decoder and only use the EGATE encoder efficiently generate compact and informative embeddings for previously unseen H-graphs, enabling downstream tasks such as classification, optimization, or quantum model parameterization.

The architecture of EGATE and its components are illustrated in Fig.~\ref{fig:model}. EGATE adopts an encoder–decoder structure to learn a graph-level representation, denoted by $\vec{g}$.
The encoder begins with the graph neural network (GNN) block that aggregates information across the H-graph from message passing. During this aggregation, structural descriptors and physical parameters, such as interaction strengths, are synthesized into higher-level representations. To implement this block, we adapt the edge-featured graph attention network (EGAT) architecture originally proposed in Ref.~\cite{EGAT}, introducing modifications tailored to our problem setting.

The GNN block starts by feeding the original node and edge descriptors, $\textbf{O}^0 =\textbf{O} =  [\vec{o}^{\,0}_i]_{i=1}^n$ and $\textbf{E}^0 =\textbf{E} =   [ \vec{e}^{\,0}_{ij}]_{\langle i,j \rangle}$, into the first EGAT layer. A total of $L$ EGAT layers are then applied sequentially. The $l$th EGAT layer takes the representations produced by its predecessor layer, $(\textbf{O}^{l-1}, \textbf{E}^{l-1})$, and emits refined descriptors, $(\textbf{O}^l, \textbf{E}^l)$. 
Within each EGAT layer, the edge module is applied after the node module. The node module produces higher-level node representations via an edge-integrated attention mechanism, incorporating edge features into both the message computation and the attention coefficients. The edge module mirrors the same attention scheme under a node-transit strategy, after which the resulting features are passed through a multi-layer perceptron (MLP) to obtain the higher-level edge representations. 
A schematic overview of the node and edge modules is provided in Fig.~\ref{fig:model}, and detailed pseudocode is given in and Algs.~\ref{appendix:alg:node module} and~\ref{appendix:alg:edge module}. 

The GNN block produces final node and edge representations, denoted as $\textbf{O}^{\mathrm{fin}}$ and $\textbf{E}^{\mathrm{fin}}$, by applying a merge layer. One option is to aggregate the outputs of all $L$ EGAT layers via concatenation, yielding hierarchical features as 
\begin{equation}
    \vec{o}_{i}^{\,\mathrm{fin}} = \mathop{\big\|}_{l=1}^{L} \vec{o}_{i}^{\, l},\quad
    \vec{e}_{ij}^{\,\mathrm{fin}} = \mathop{\big\|}_{l=1}^{L} \vec{e}_{ij}^{\, l},
    \label{eq:concat merge layer}
\end{equation}
where $\mathop{\|}$ denotes concatenation operator. As a result, the final node and edge representations have shapes $O^{\mathrm{fin}}\in\mathbb{R}^{n\times Ld_o}$ and $E^{\mathrm{fin}}\in\mathbb{R}^{m\times Ld_e}$, respectively. While the concatenation merge layer retains all layer-specific information, it may lead to a substantial memory burden for large H-graphs. A lightweight alternative is to aggregate the layer outputs by an element-wise mean, 
\begin{equation}
    \vec{o}_{i}^{\,\mathrm{fin}} = \frac{1}{L} \sum_{l=1}^L \vec{o}_{i}^{\,l},\quad
    \vec{e}_{ij}^{\,\mathrm{fin}} = \frac{1}{L} \sum_{l=1}^L \vec{e}_{ij}^{ \,l},
    \label{eq:mean merge layer}
\end{equation}
which significantly reduces memory usage at the expense of reduced access to layer-specific information.

The second part of the encoder is a pooling block that performs dimensionality reduction on the encoded graph representations after the EGAT layers. 
One of the simplest non-trainable approaches is sum pooling, which aggregates final node and edge embeddings separately by summation. Under sum pooling, the latent vector $\vec{g}$ is given by
\begin{equation}
    \vec{g} = \left[\sum_{i=1}^n \vec{o}^{\,\mathrm{fin}}_i \mathbin{\bigg\Vert} \sum_{\langle i,j\rangle}^m\vec{e}_{ij}^{\,\mathrm{fin}}\right],
    \label{eq:sum pooling}
\end{equation}
where $\vec{g} \in \mathbb{R}^{L(d_o+d_e)}$ for the concatenation merge layer and $\vec{g} \in\mathbb{R}^{d_o+d_e}$ for the element-wise mean merge layer.
To obtain a more expressive latent vector $\vec{g}\in\mathbb{R}^d$ with a flexible dimension, we also use a trainable pooling operation implemented as an MLP-based projection, which flattens the final node and edge feature matrix and projects it to an embedding of dimension $d$. One can also consider other various pooling strategies for graph embedding or graph representation learning~\cite{liu2023graphpoolinggraphneural,li2024graph,liu2025graph}.




The final component of EGATE is a decoder that reconstructs the original H-graph from $\vec{g}$. Since the input H-graph contains multi-dimensional node and edge features, the decoder must be designed to  reconstruct both types of information simultaneously. 
Accordingly, we train EGATE using a composite reconstruction loss that combines node- and edge-level errors. It is important to note a potential pitfall common to autoencoder-based architectures: a decoder with overly excessive expressivity, such as a deep MLP, can achieve low reconstruction loss regardless of the quality of the latent representation. This undermines the encoder's role and compromises the model’s ability to extract meaningful structure from the data. To mitigate this issue and encourage the encoder to capture meaningful structural embeddings, we deliberately restrict the expressivity of the decoder.

\begin{figure*}[t]
\centering
\includegraphics[width=0.8\textwidth]{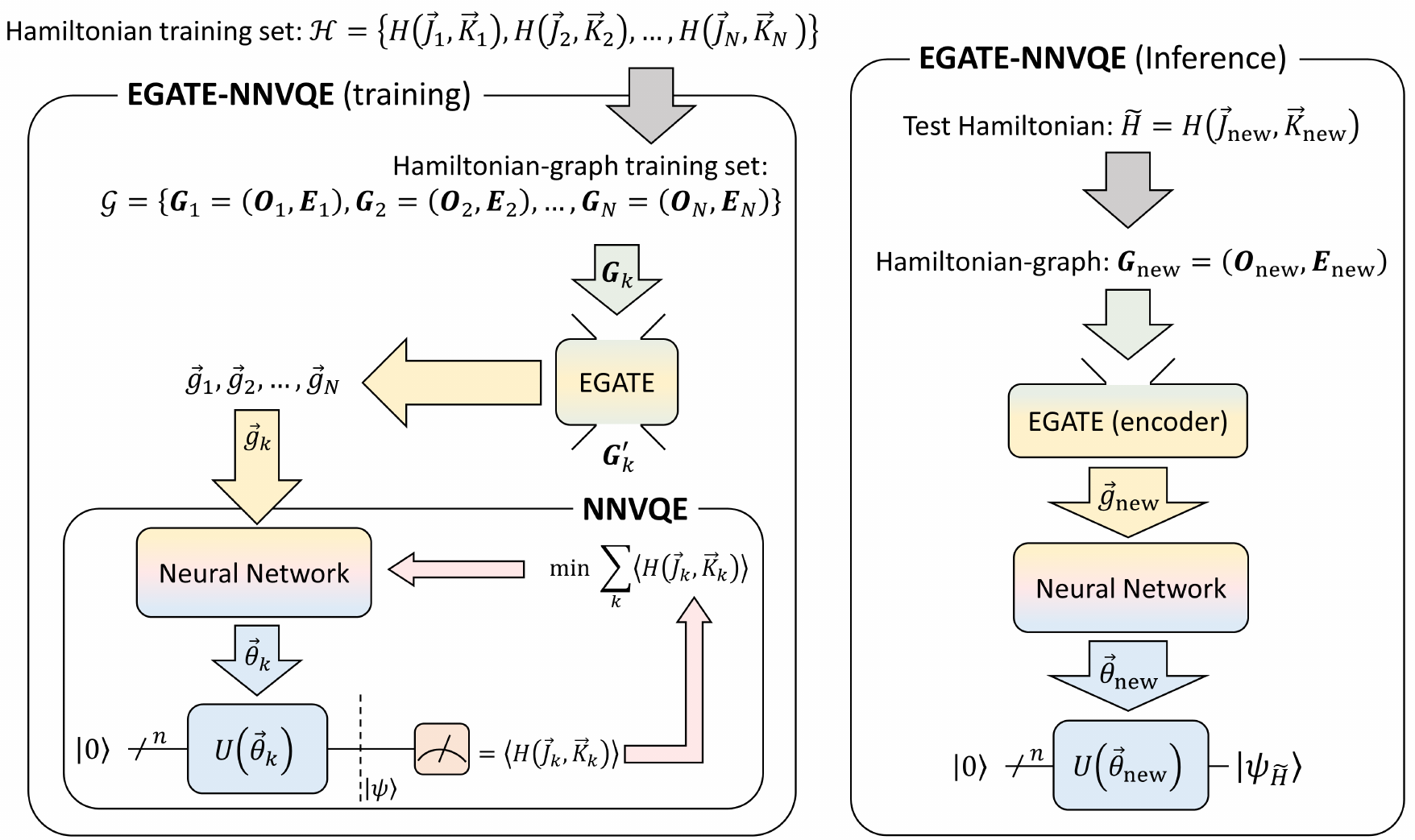} 
\caption{Schematic overview of EGATE-NNVQE for training (left) and inference (right).}
\label{fig:EGATE-NNVQE workflow}
\end{figure*}

\section{Overall Workflow}
In this section, we describe the training and testing workflow of EGATE-NNVQE, illustrated in Fig.~\ref{fig:EGATE-NNVQE workflow}. 
Let $\mathcal{H}=\lbrace H(\vec{J}_k, \vec{K}_k)\rbrace_{k=1}^N$ denotes a set of training Hamiltonians, where $\vec{J}_k$ and $\vec{K}_k$ are the tunable parameter vectors in Eq.~(\ref{eq:general two-local H}).

As we described in Sec.~\ref{sec:NNVQE}, NNVQE in Ref.~\cite{NN-VQA} trains its classical NN with $(\vec{J}_k,\vec{K}_k)$ as input. In contrast, EGATE-NNVQE takes the latent vector $\vec{g}_k$ as input to the NNVQE component as illustrated in Fig.~\ref{fig:EGATE-NNVQE workflow}.
To obtain $\vec{g}_k$, we first map each $H(\vec{J}_k,\vec{K}_k)$ to the corresponding H-graph $\mathbf{G}_k=(\mathbf{O}_k,\mathbf{E}_k)$. We then train EGATE on the graph dataset $\mathcal{G}=\lbrace\mathbf{G}_k\rbrace_{k=1}^{N}$ by minimizing a reconstruction loss $\mathcal{L}_{\mathrm{rec}}$, producing latent vectors $\lbrace\vec{g}_k\rbrace_{k=1}^{N}$ that encode the structural and parametric information of the Hamiltonians. Throughout this paper, we use
\begin{equation}
\mathcal{L}_{\mathrm{rec}} := \mathrm{MSE}(\mathbf{O},\mathbf{O}')+\beta\cdot\mathrm{MSE}(\mathbf{E},\mathbf{E}'),    
\end{equation}
where $\beta\ge 0$ controls the relative weight of the edge reconstruction term ($\beta=1$ unless stated otherwise). Given $\{\vec{g}_k\}$, the NNVQE component in EGATE-NNVQE is trained in the same way as in the original NNVQE, except that its input is replaced by $\vec{g}_k$.  Importantly, EGATE and NNVQE are trained separately using the same set of training Hamiltonians.

Inference on a new Hamiltonian $\tilde{H}=H(\vec{J}_{\mathrm{new}},\vec{K}_{\mathrm{new}})$, defined by a previously unseen parameter configuration $(\vec{J}_{\mathrm{new}}, \vec{K}_{\mathrm{new}})$, proceeds as shown on the right side of the Fig.~\ref{fig:EGATE-NNVQE workflow}. 
We first construct H-graph $\mathbf{G}_{\mathrm{new}}$ of $\tilde{H}$ and obtain $\vec{g}_{\mathrm{new}}$ using the trained EGATE encoder. 
The trained NN then directly outputs circuit parameters $\vec{\theta}_{\mathrm{new}}=f_{\phi^{*}}(\vec{g}_{\mathrm{new}})$ without additional optimization, where $\phi^{*}$ denotes the trained NN weights. These parameters define a initial state $|\psi_{\tilde{H}}\rangle=U(\vec{\theta}_{\mathrm{new}})|0\rangle^{\otimes n}$, which can be used as an initial (reference) state for downstream quantum algorithms with rigorous convergence guarantees, such as QPE and QSMs.


\section{Result}
\label{sec:result}
In this section, we demonstrate improved generalization performance and mitigation of the BP phenomenon for EGATE-NNVQE relative to the NNVQE baseline, which denotes the model without graph-based representation learning. As a concrete demonstration, we consider four Hamiltonian families: 1D XXZ Heisenberg spin chain without an external field $(H_{\text{XXZ}})$ and with a transverse field $(H_{\text{XXZ+X}})$, as well as two-dimensional (2D) XXZ and XYZ Hamiltonians defined on a $3\times3$ square lattice with 9 qubits, denoted by $H_{\mathrm{XXZ}}^{3\times3}$ and $H_{\mathrm{XYZ}}^{3\times3}$, respectively. The 1D XXZ Hamiltonian is given by
\begin{equation}
    H_{\text{XXZ}} = \sum_{i=1}^{n}\left(\sigma^x_i\sigma^x_{i+1} + \sigma^y_i\sigma^y_{i+1} +J^{zz} \sigma^z_i\sigma^z_{i+1}\right) ,
    \label{eq:XXZ} 
\end{equation}
and the XXZ model with a transverse field is
\begin{align}
    H_{\text{XXZ+X}} = &\sum_{i=1}^{n}\left(\sigma^x_i\sigma^x_{i+1} + \sigma^y_i\sigma^y_{i+1} +J^{zz} \sigma^z_i\sigma^z_{i+1}\right)\notag\\
    &+\sum_{i=1}^n K^x \sigma^x_i.
    \label{eq:XXZ+X}
\end{align}
For the 2D lattice, the XXZ Hamiltonian is defined as
\begin{align}
    H_{\mathrm{XXZ}}^{3\times3}
    = &\sum_{\langle i,j \rangle_1}
        \left(
            \sigma^x_i \sigma^x_j
          + \sigma^y_i \sigma^y_j
          + J^{zz}_1\sigma^z_i \sigma^z_j
        \right)\notag\\
    & + \sum_{\langle i,j \rangle_2}
        \left(
            \sigma^x_i \sigma^x_j
          + \sigma^y_i \sigma^y_j
          + J^{zz}_2\sigma^z_i \sigma^z_j
        \right),
    \label{eq:2D_XXZ}
\end{align}
while the corresponding XYZ Hamiltonian is
\begin{align}
    H_{\mathrm{XYZ}}^{3\times3}
    =& \sum_{\langle i,j \rangle_1}
        \left(
            \sigma^x_i \sigma^x_j
          +J^{yy} \sigma^y_i \sigma^y_j
          + J^{zz}_1\sigma^z_i \sigma^z_j
        \right)\notag\\
    & + \sum_{\langle i,j \rangle_2}
        \left(
            \sigma^x_i \sigma^x_j
          +J^{yy} \sigma^y_i \sigma^y_j
          + J^{zz}_2\sigma^z_i \sigma^z_j
        \right).
    \label{eq:2D_XYZ}
\end{align}
Here, $J$ and $K$ denote tunable coupling parameters corresponding to two-local and one-local interaction strengths, respectively. For the 1D Hamiltonians, we impose periodic boundary conditions, such that site indices are taken modulo $n$, i.e., $(i=n)+1 = 1$. For 2D Hamiltonians on the $3\times 3$ lattice, $\langle i,j\rangle_1$ and $\langle i,j\rangle_2$ denote the sets of nearest-neighbor and next-nearest-neighbor edges, respectively, with $i<j$.

In the original NNVQE benchmark~\cite{NN-VQA}, the XXZ chain with a uniform longitudinal field, $H_{\mathrm{XXZ+Z}}$ (Eq.~(\ref{eq:XXZ spin})), was considered. In this setting, the Hamiltonian preserves the U(1) symmetry associated with the conservation of total magnetization $S_{\mathrm{tot}}^{z}$, i.e., $[H, S_{\mathrm{tot}}^{z}]=0$, which renders the learning task relatively tractable.
By contrast, introducing a transverse field breaks this U(1) symmetry. We therefore benchmark our method on $H_{\mathrm{XXZ+X}}$ to assess generalization performance in a symmetry-breaking setting that goes beyond the scope of prior NNVQE studies.

\begin{figure*}
\centering
\includegraphics[width=\textwidth]{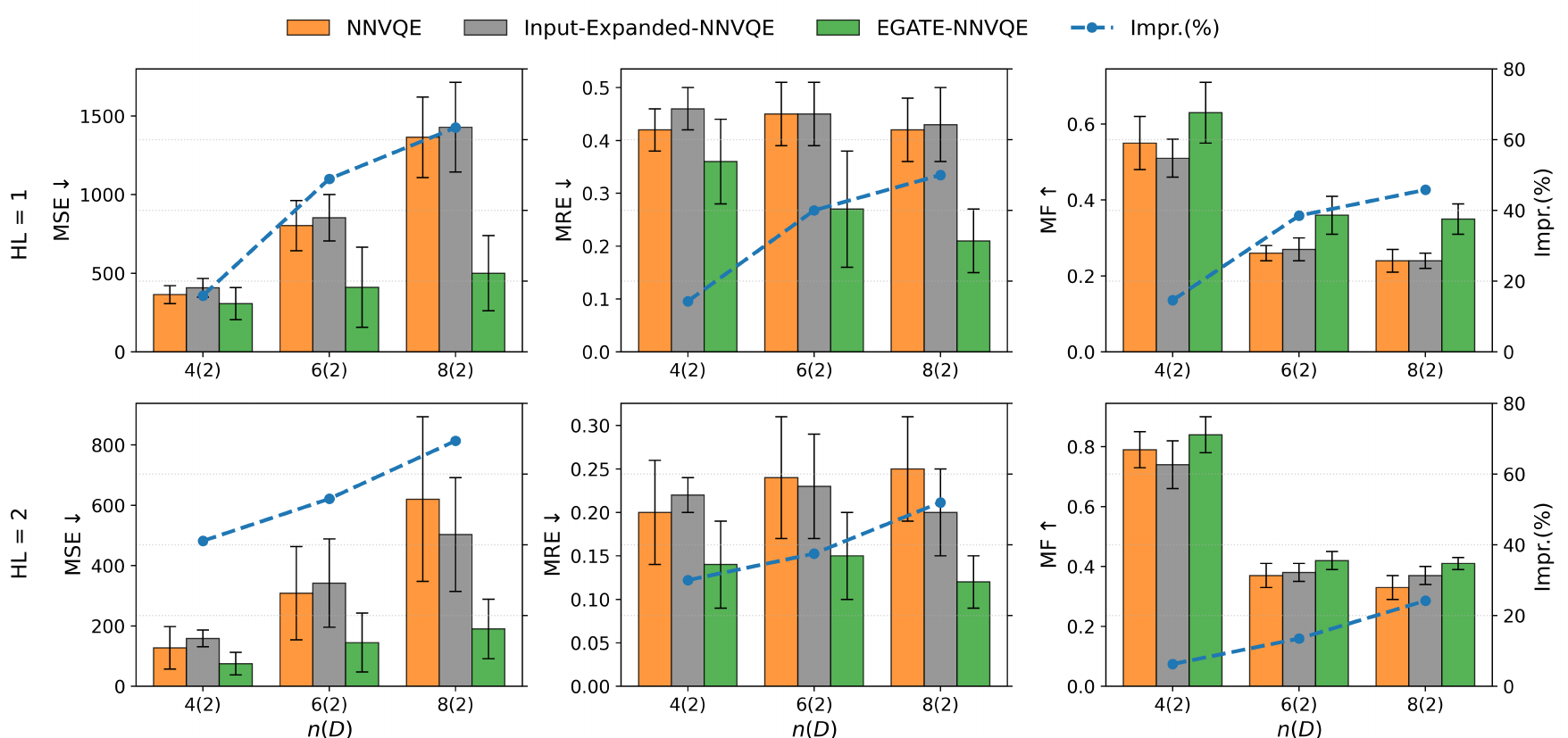}\\
(a) $H_\text{XXZ}$\\
\includegraphics[width=\textwidth]{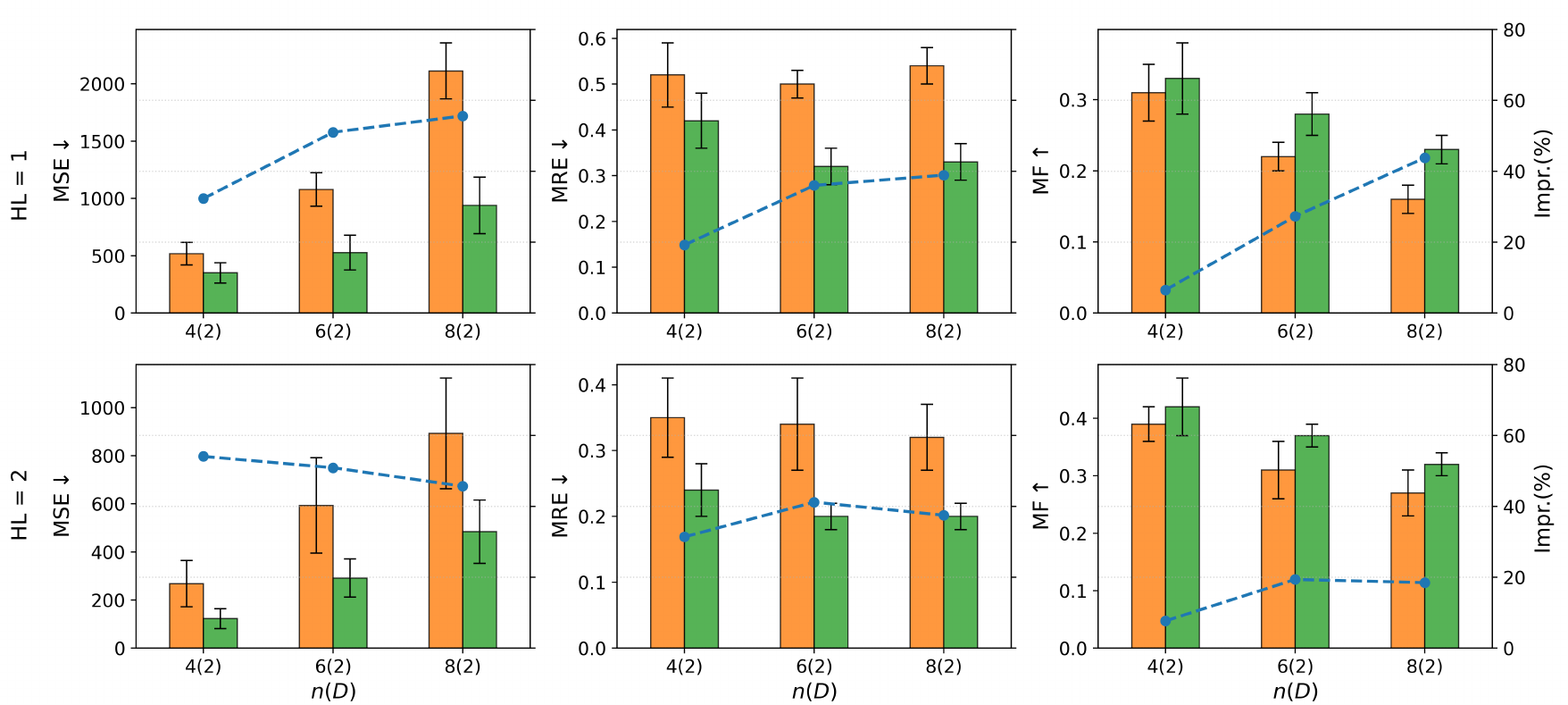}\\
(b) $H_\text{XXZ+X}$\\
\caption{Generalization performance of NNVQE baseline and EGATE-NNVQE on 1D Hamiltonian families, $H_{\text{XXZ}}$ (Eq.~(\ref{eq:XXZ})) and $H_{\text{XXZ+X}}$ (Eq.~(\ref{eq:XXZ+X})), with system size $n\in{\lbrace4,6,8\rbrace}$ qubits, where $D$ denotes the number of ansatz blocks. Performance is evaluated using MSE and MRE (lower is better, $\downarrow$), and MF (higher is better, $\uparrow$). Rows correspond to the number of hidden layers (HL = 1 or 2) in the NNVQE. Bars indicate mean $\pm$ standard deviation over 10 random seeds, where each seed’s value is the average over all test instances. In (a), each seed uses 20 training and 1,000 test instances of $H_\text{XXZ}$; in (b), 400 training and 40,000 test instances of $H_\text{XXZ+X}$. 
The right axis (Impr.) reports the relative improvement (\%) of EGATE-NNVQE (green) over the NNVQE (orange) baseline, shown as a blue dashed line. EGATE-NNVQE consistently improves all metrics across settings, with relative improvements that remain comparable or increase as the system size ($n$) grows, indicating that the NNVQE baseline increasingly struggles to capture relevant structural information at larger scales. To isolate the role of the EGATE representation, (a) includes input-expanded-NNVQE (gray), where the NNVQE input dimension is matched to the EGATE latent size. Its performance is comparable to, or worse than, that of the NNVQE baseline, indicating that the observed improvements arise from the structured latent representation learned by EGATE rather than from increased dimensionality or model capacity.}

\label{fig:1D_generalization_result}
\end{figure*}

\begin{figure*}[t]
\centering

\includegraphics[width=\textwidth]{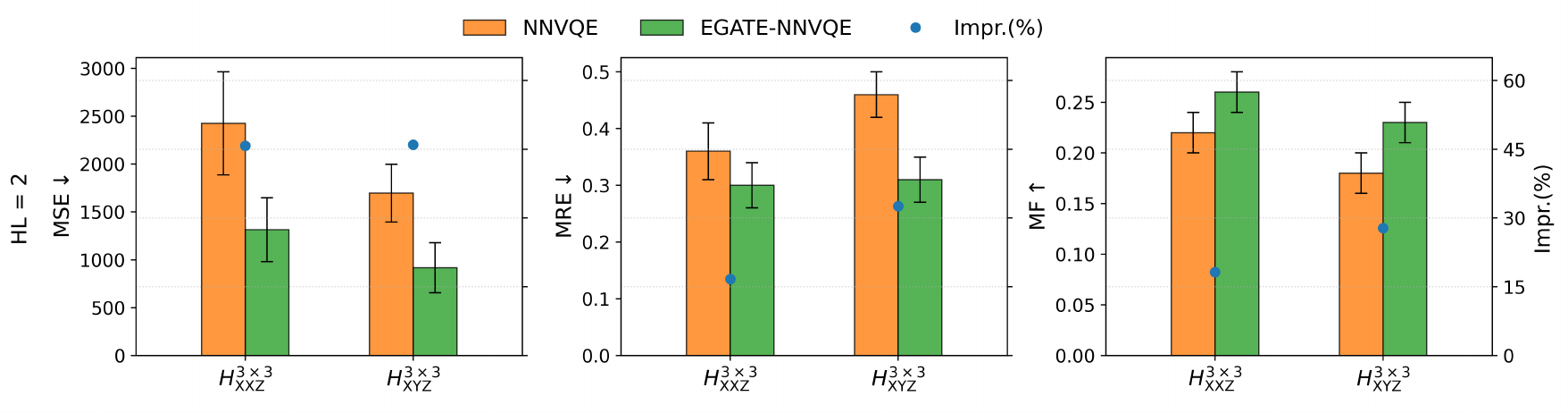}

\caption{
Generalization performance of NNVQE baseline and EGATE-NNVQE on 2D Hamiltonians on $3\times3$ lattices, $H_{\mathrm{XXZ}}^{3\times3}$ (Eq.~(\ref{eq:2D_XXZ})) and $H_{\mathrm{XYZ}}^{3\times3}$ (Eq.~\ref{eq:2D_XYZ})). The same performance metrics and notation as in Fig.~\ref{fig:1D_generalization_result} are used. In contrast to the one-dimensional case, the system size is fixed ($n=9$, $D=2$), and the horizontal axis indexes the Hamiltonian family. Results are reported as mean $\pm$ standard deviation over 10 random seeds, where each seed’s value is averaged over all test instances. We use 400 training and 40,000 test instances for $H_{\mathrm{XXZ}}^{3\times3}$, and 405 training and 32,000 test instances for $H_{\mathrm{XYZ}}^{3\times3}$. Consistent with Fig.~\ref{fig:1D_generalization_result}, EGATE-NNVQE outperforms NNVQE baseline across all metrics for both Hamiltonian families.}
\label{fig:2D_generalization_result}
\end{figure*}

\subsection{Generalization}
\label{sec:generalization}
Generalization performance is evaluated using three metrics: mean squared error (MSE), mean relative energy error (MRE), and mean fidelity (MF), where $\text{MF}\in[0,1]$. The MSE and MRE are computed between the predicted and exact ground‑state energies, while the MF is evaluated between the predicted and exact ground states. For each metric, values are first averaged over all test Hamiltonian instances. We then report the mean $\pm$ standard deviation (SD) over 10 independent runs corresponding to different random initializations of the NN weights. The initial weights are sampled from a normal distribution $\mathcal{N}(0.0,0.1)$. Finally, we validate the generalization advantage of our method by evaluating the relative improvement of EGATE-NNVQE with respect to the NNVQE baseline for each performance metric. Since lower values indicate better performance for MSE and MRE, while higher values indicate better performance for MF, we define the relative improvement with an appropriate sign convention. For instance, we use
\begin{equation}
    \frac{\mathrm{MSE}_\mathrm{NNVQE}-\mathrm{MSE}_\mathrm{EGATE-NNVQE}}{\mathrm{MSE}_\mathrm{NNVQE}}\times 100\;(\%)
\end{equation}
and
\begin{equation}
    \frac{\mathrm{MF}_\mathrm{EGATE-NNVQE}-\mathrm{MF}_\mathrm{NNVQE}}{\mathrm{MF}_\mathrm{NNVQE}}\times 100\;(\%).
\end{equation}
The same sign convention as for MSE is applied to MRE.

We adopt the ladder-wise hardware-efficient ansatz (HEA)~\cite{HEA,NN-VQA}, where $D$ denotes the number of the ansatz block; details of the ansatz structure are provided in~\ref{appendix:sec:ansatz}. The classical NN in NNVQE is implemented as an MLP, and we consider architectures with one or two hidden layers (HL = 1 or 2).
The NNVQE component in EGATE-NNVQE (i.e. NN and ansatz) is architecturally identical to the NNVQE baseline in each setting (HL = 1 or 2), using the same hidden layer units. The only differences are the input representation and input dimensionality. Complete hyperparameter settings for NN are summarized in~\ref{appendix:sec:NN structure}.


For $H_{\mathrm{XXZ}}$ (Eq.~(\ref{eq:XXZ})), we train on 20 uniformly spaced values of $J^{zz}\in[-3,3]$ and evaluate on 1,000 uniformly spaced points in $[-10,10]$ for each system size. Since $J^{zz}$ is the only tunable parameter, the NNVQE baseline always takes a one-dimensional input, $J^{zz}$. We encode each Hamiltonian as an H-graph whose node features are one-hot vectors $\vec{o}_i=\mathbf{e}_i\in\mathbb{R}^n$, and whose edge features are $\vec{e}_{ij}=(J_{ij}^{xx},J_{ij}^{yy},J_{ij}^{zz})=(1,1,J^{zz})$. EGATE uses mean merge (Eq.~(\ref{eq:mean merge layer})) and sum pooling (Eq.~(\ref{eq:sum pooling})), with a single-hidden-layer perceptron as the decoder, unless stated otherwise. In this setting, EGATE produces latent representations of dimension 7, 9, and 11 for 4-, 6-, and 8-qubit systems, respectively. 
For $H_{\mathrm{XXZ}+X}$ (Eq.~(\ref{eq:XXZ+X})), we additionally include a one-local transverse-field term $K^x$, modeled as a self-loop in the H-graph and encoded by concatenating $K^{x}$ to the node features. We use permutation-invariant node features $\vec{o}_i=(1,K^x)$ and the same edge features $\vec{e}_{ij}=(1,1,J^{zz})$. Then, EGATE produces a fixed 5-dimensional latent vector independent of $n$, while the NNVQE baseline takes a two-dimensional input $(J^{zz},K^x)$. We sample 20 uniformly spaced values in $[-3,3]$ for each parameter for training and 200 values in $[-10,10]$ for each parameter for testing, yielding 400 training and 40,000 test Hamiltonians.

For 2D Heisenberg models on $3\times 3$ lattice, $H_{\mathrm{XXZ}}^{3\times 3}$ (Eq.~(\ref{eq:2D_XXZ})) and $H_{\mathrm{XYZ}}^{3\times 3}$ (Eq.~(\ref{eq:2D_XYZ})), which share the same lattice geometry and do not contain one-local terms, we use the lattice coordinates as node features, $\vec{o}_i=(x_i,y_i)$ with $x_i,y_i\in\{-1,0,1\}$.
We distinguish nearest- and next-nearest-neighbor couplings by assigning edge features $\vec{e}_{ij}=(J_{ij}^{xx},J_{ij}^{yy},J_{ij}^{zz})$ as $(1,1,J^{zz}_k)$ for $H_{\mathrm{XXZ}}^{3\times 3}$ and $(1,J^{yy},J^{zz}_k)$ for $H_{\mathrm{XYZ}}^{3\times 3}$, where $k=1$ for nearest-neighbor edges and $k=2$ for next-nearest-neighbor edges.
For $H_{\mathrm{XXZ}}^{3\times 3}$, the NNVQE baseline takes a two-dimensional input $(J^{zz}_1,J^{zz}_2)$, whereas EGATE-NNVQE conditions the predictor on an 8-dimensional latent vector produced by EGATE. We sample $J^{zz}_1$ and $J^{zz}_2$ on a 20-point uniform grid per parameter over $[-3,3]$ for training and on a 200-point grid over $[-10,10]$ for testing, yielding 400 training instances and 40,000 test instances.
For $H_{\mathrm{XYZ}}^{3\times 3}$, the NNVQE baseline takes a three-dimensional input $(J^{yy},J^{zz}_1,J^{zz}_2)$, while EGATE-NNVQE again uses an 8-dimensional latent representation. Training uses a 5-point uniform grid for $J^{yy}\in[-1,1]$ and 9-point uniform grids for $J^{zz}_1,J^{zz}_2\in [-2,2]$, yielding 405 training instances, while testing employs 20- and 40-point uniform grids over $[-4,4]$ and $[-7,7]$ for $J^{yy}$ and $(J^{zz}_1,J^{zz}_2)$, respectively, resulting in 32,000 test instances.
 Additional experimental details, including training procedures and hyperparameter selection, are provided in Appendix~\ref{appendix:sec:experimental setting}. 

Figure~\ref{fig:1D_generalization_result} compares the generalization performance of NNVQE and EGATE-NNVQE on 1D Heisenberg models over 10 random seeds, for (a) $H_{\mathrm{XXZ}}$, and (b) $H_{\mathrm{XXZ+X}}$. Bars show mean $\pm$ SD of MSE, MRE, and MF (lower is better for MSE and MRE, while higher is better for MF). 
Orange and green bars correspond to NNVQE baseline and EGATE-NNVQE, respectively. 
The blue dashed curve (right axis) indicates the relative improvement of EGATE-NNVQE over the NNVQE baseline. Across both Hamiltonian families and all settings we consider, EGATE-NNVQE (green bar) consistently outperforms the NNVQE baseline (orange bar) in every metric. In particular, the relative improvement (blue dashed line) generally becomes more pronounced as the system size $n$ increases, suggesting that the NNVQE baseline increasingly struggles to capture relevant Hamiltonian structure at larger scales.

To assess whether these gains arise merely from increased input dimensionality, we introduce an input-expanded NNVQE baseline (gray bar) in Fig.~\ref{fig:1D_generalization_result}(a), in which the scalar coupling input is replicated to match the latent dimension used by EGATE-NNVQE (e.g., replacing $J^{zz}$ with $(J^{zz},\dots,J^{zz})$). This baseline matches the predictor architecture, input dimension, and parameter count of EGATE-NNVQE, differing only in the input representation. The input-expanded-NNVQE fails to close the performance gap and is often comparable to or worse than the NNVQE baseline, demonstrating that the observed improvements stem from the informativeness of EGATE’s latent representation, rather than from increased dimensionality alone. 

By contrast, the relative improvement curve no longer increases monotonically with $n$ and instead exhibits mild saturation or a slight decline in panel (b) (HL$=2$). This behavior is consistent with a representational bottleneck: in our $H_{\mathrm{XXZ+X}}$ setting, EGATE uses a fixed-dimensional latent vector as the system size grows. These results suggest that scaling the latent dimension with problem size may be beneficial. However, as demonstrated by the input-expanded baseline, increasing dimensionality alone is insufficient unless the representation captures meaningful Hamiltonian structure.

Figure~\ref{fig:2D_generalization_result} compares the generalization performance of NNVQE and EGATE-NNVQE for 2D Heisenberg models on $3\times 3$ lattice, $H_{\mathrm{XXZ}}^{3\times 3}$ and $H_{\mathrm{XYZ}}^{3\times 3}$. Overall, even on the $3\times 3$ lattice 2D Hamiltonian families, EGATE-NNVQE preserves the advantages observed in the 1D setting, demonstrating robust generalization gains over the NNVQE baseline on both dimensionality and interaction structure.

\subsubsection{Application to Quantum Subspace Methods}
\label{Sec:SKQD}
\begin{figure*}[ht]
\centering
\begin{tabular}{c c}
    \includegraphics[width=0.42\textwidth]{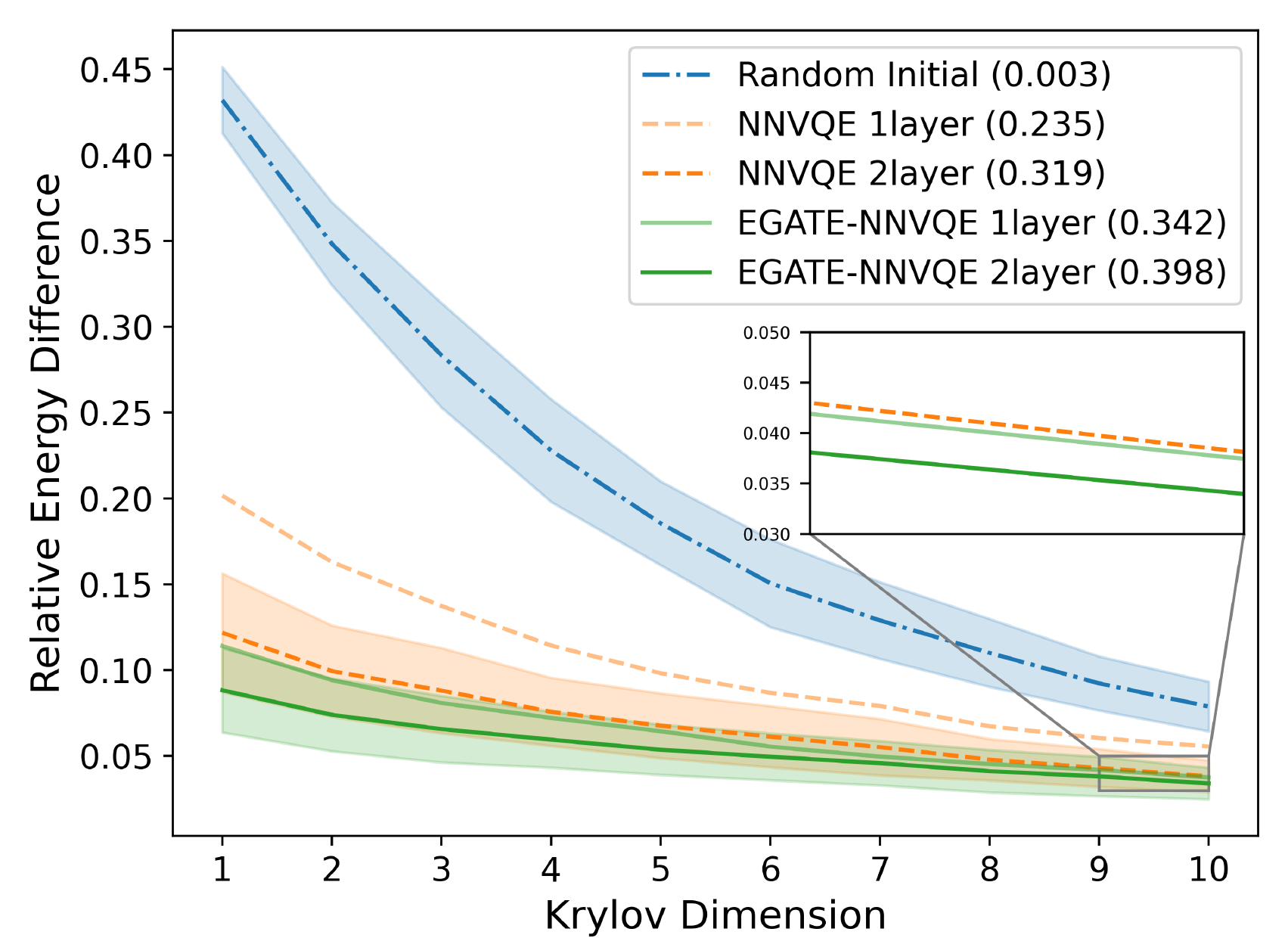} &
    \includegraphics[width=0.42\textwidth]{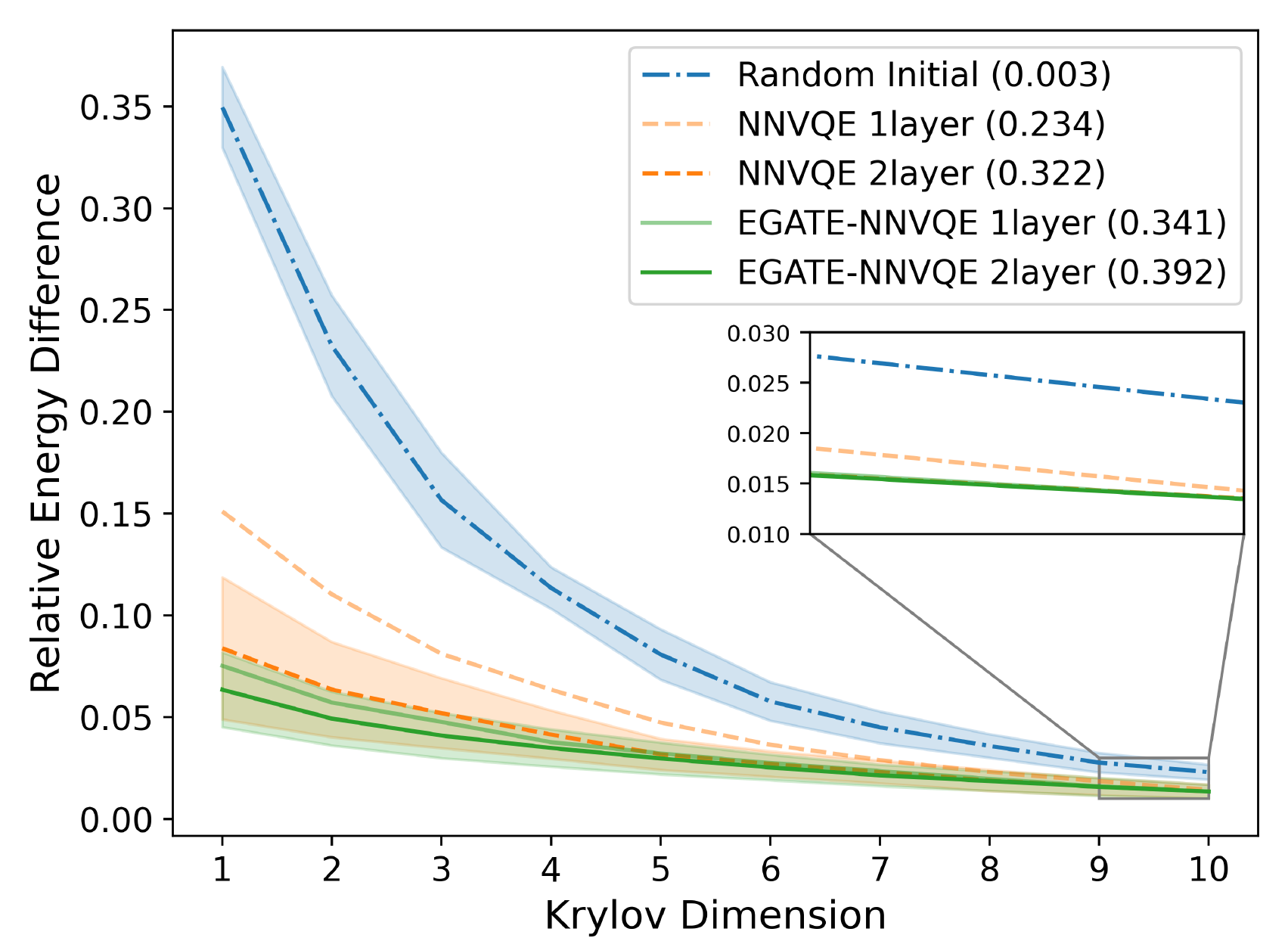} \\
    (a) $H_\text{XXZ}$ 25 shots & (b) $H_\text{XXZ}$ 50 shots\\
    
    \includegraphics[width=0.42\textwidth]{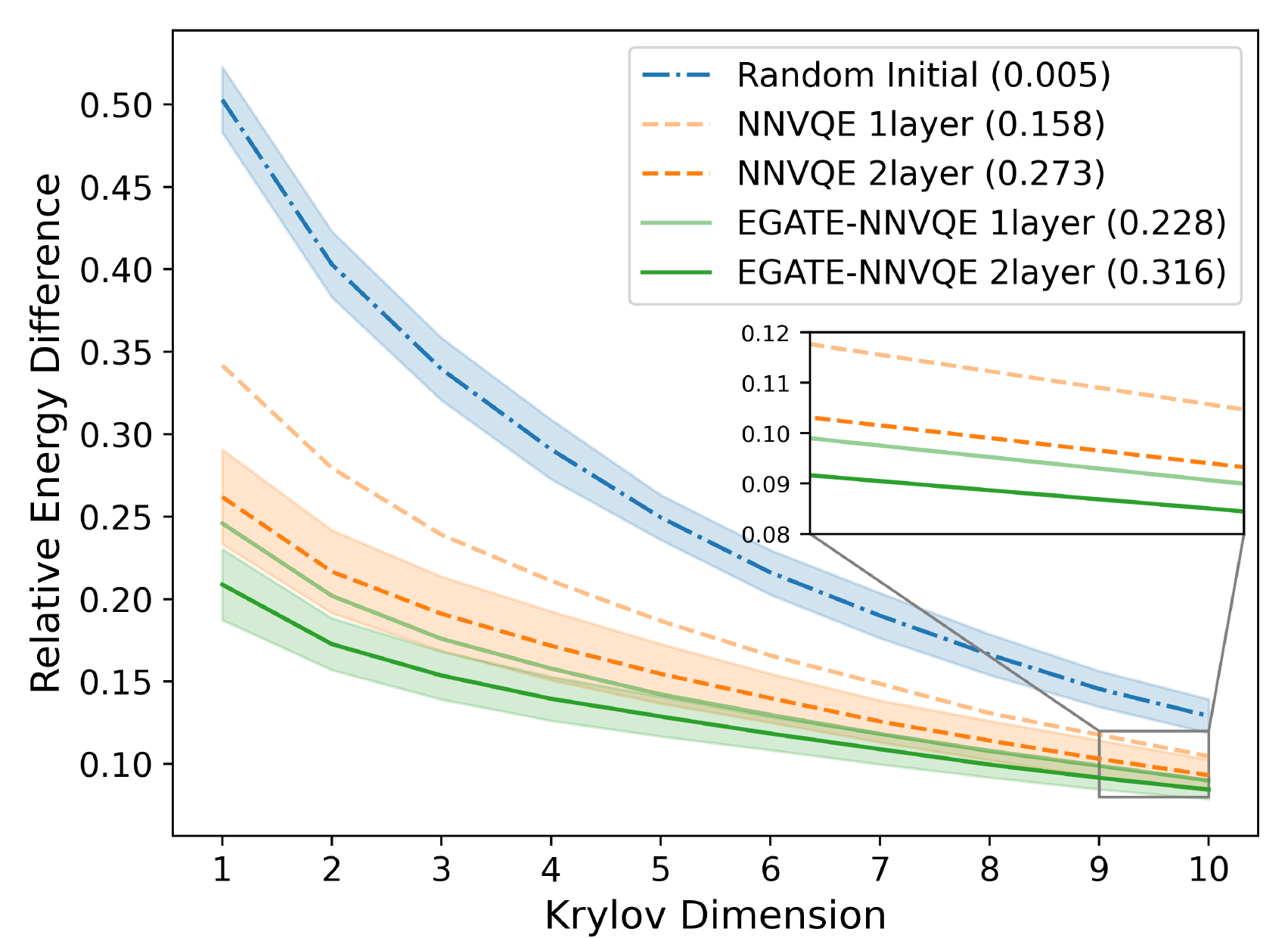} &
    \includegraphics[width=0.42\textwidth]{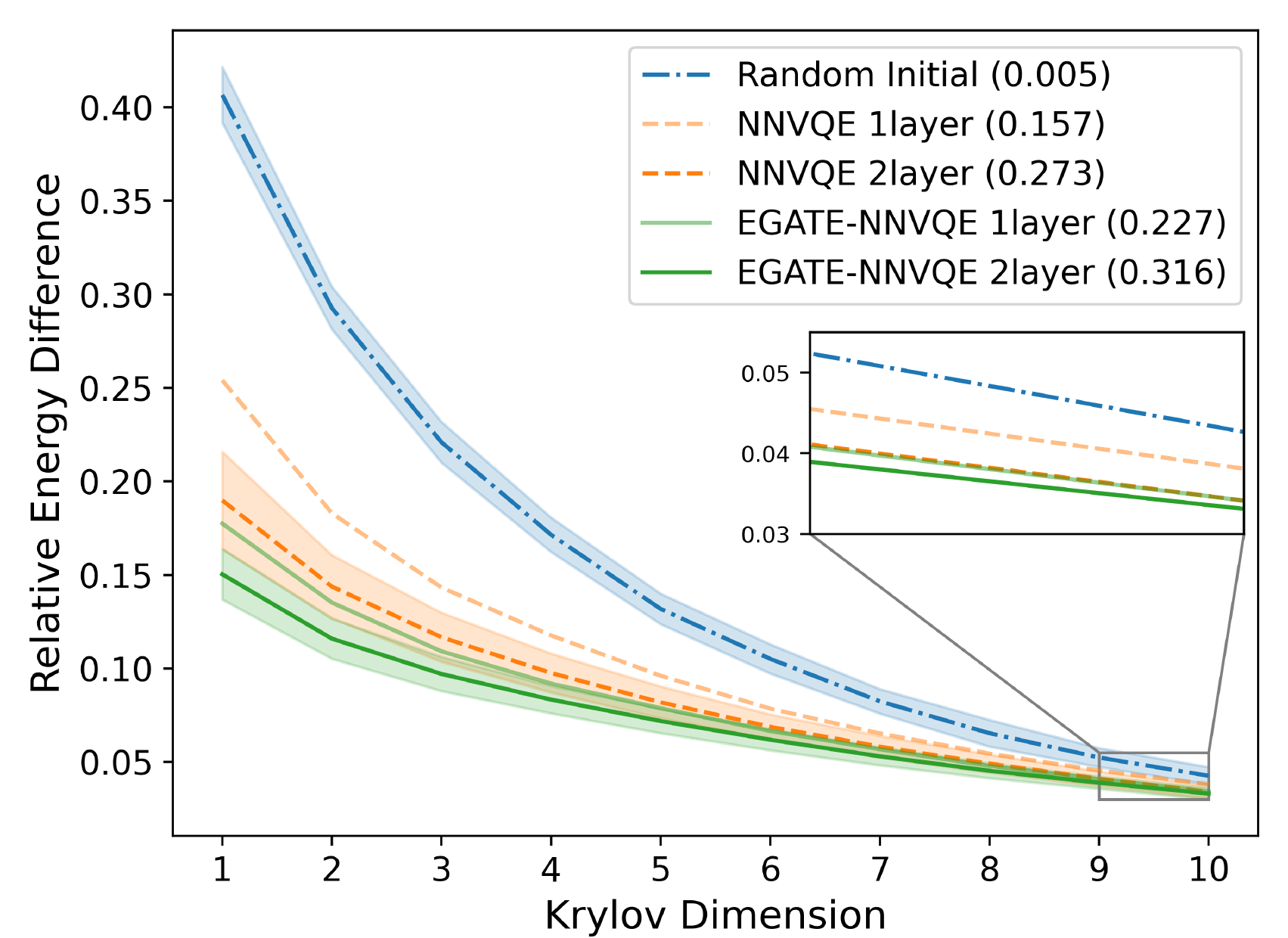} \\
    (c) $H_\text{XXZ+X}$ 25 shots & (d) $H_\text{XXZ+X}$ 50 shots\\
    
  \end{tabular}
  \caption{SKQD results using three input-state benchmarks: EGATE-NNVQE, NNVQE, and a random initial state, evaluated under noisy simulations based on an IBM fake device. Rows correspond to the 1D Hamiltonian families $H_{\mathrm{XXZ}}$ (top) and $H_{\mathrm{XXZ+X}}$ (bottom). For each case, the Krylov subspace dimension is varied up to $d=10$, and the relative energy error with respect to the exact ground-state energy is reported. Panels (a) and (b) show results for $H_{\mathrm{XXZ}}$, and panels (c) and (d) for $H_{\mathrm{XXZ+X}}$, using 25 and 50 measurement shots, respectively. For each random seed, 100 test Hamiltonians are sampled for $H_{\mathrm{XXZ}}$ (from a pool of 1,000) and 200 for $H_{\mathrm{XXZ+X}}$ (from 40,000), with results aggregated over 10 seeds. Curves show mean $\pm$ standard deviation across seeds. Values shown in parentheses in the legend denote the mean fidelity between each benchmark input state and the corresponding exact ground state. The EGATE-NNVQE and NNVQE models are those trained in the generalization experiments shown in Fig.~\ref{fig:1D_generalization_result}, and all input states are generated without additional instance-specific optimization.}
\label{fig:1D SKQD result}
\end{figure*}

\begin{figure*}[t]
\centering
\begin{tabular}{c c}
    \includegraphics[width=0.42\textwidth]{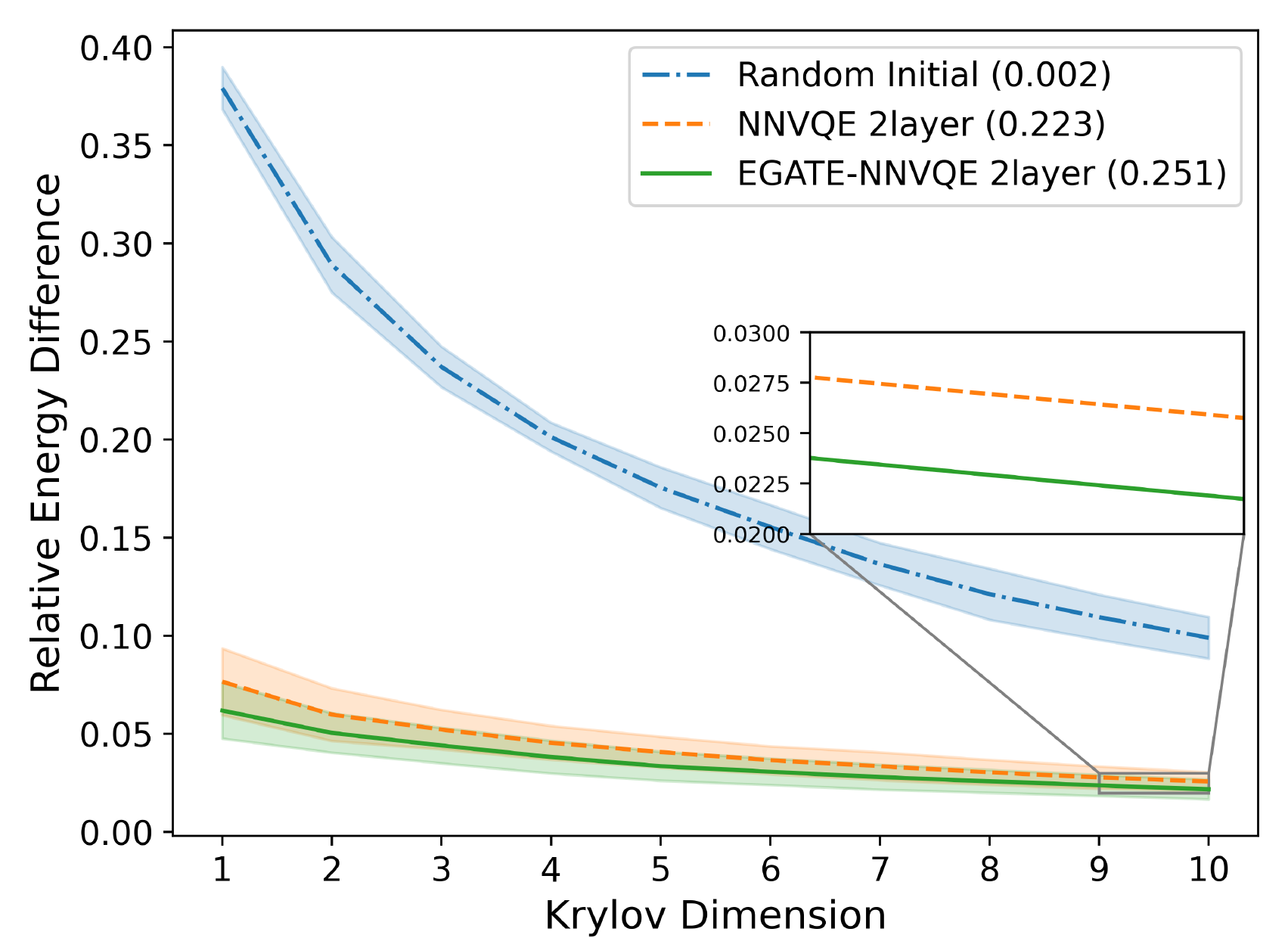} &
    \includegraphics[width=0.42\textwidth]{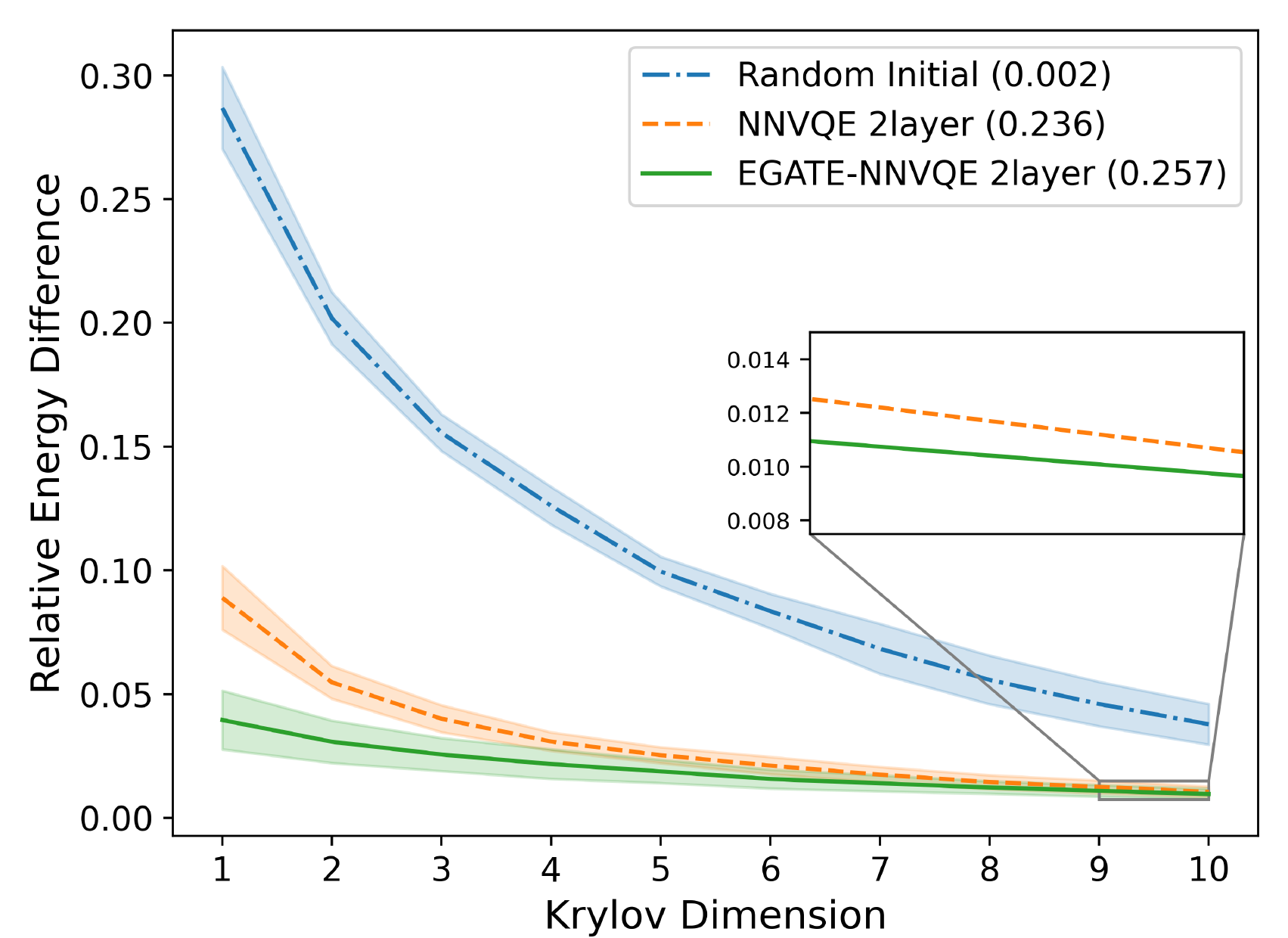} \\
    (a) $H_\text{XXZ}^{3\times3}$ 25 shots & (b) $H_\text{XXZ}^{3\times3}$ 50 shots\\
    
    \includegraphics[width=0.42\textwidth]{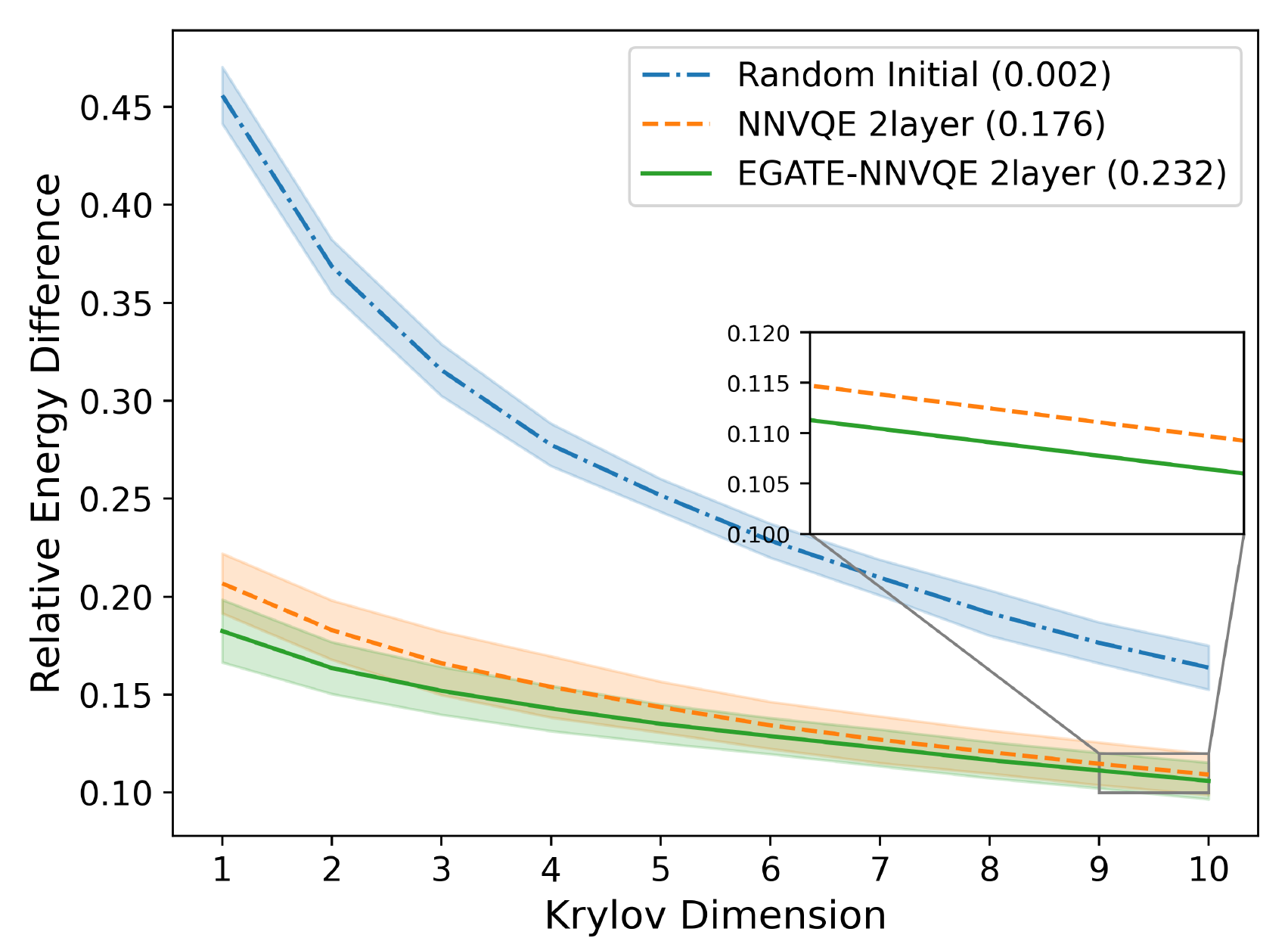} &
    \includegraphics[width=0.42\textwidth]{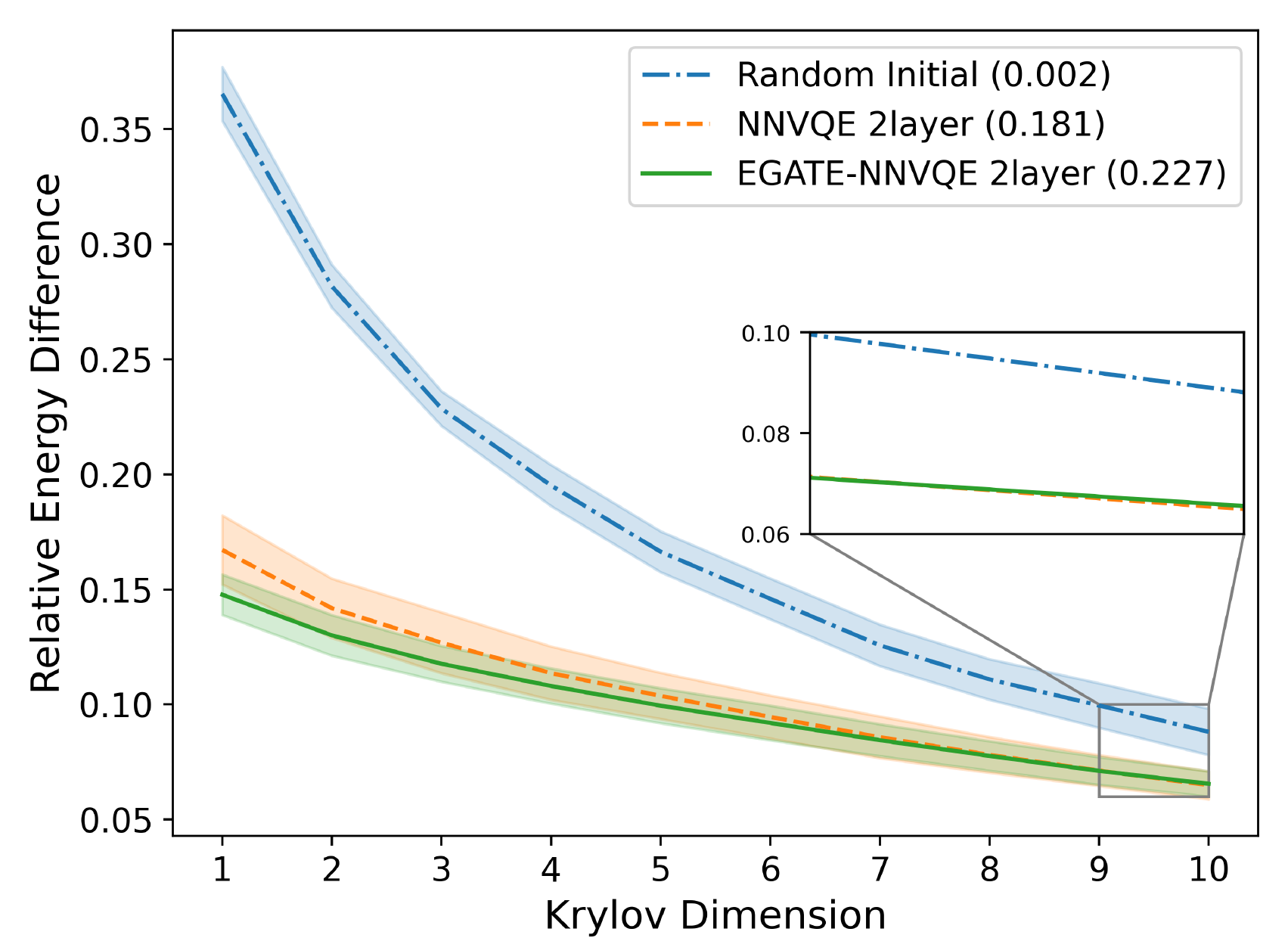} \\
    (c) $H_\text{XYZ}^{3\times3}$ 25 shots & (d) $H_\text{XYZ}^{3\times3}$ 50 shots
  \end{tabular}
  \caption{SKQD results for the 2D $3\times 3$ Heisenberg models $H_{\mathrm{XXZ}}^{3\times 3}$ and $H_{\mathrm{XYZ}}^{3\times 3}$, obtained using the same experimental settings as in Fig.~\ref{fig:1D SKQD result}. Panels (a) and (b) show results for $H_{\mathrm{XXZ}}^{3\times 3}$, while (c) and (d) show results for $H_{\mathrm{XYZ}}^{3\times 3}$, using 25 and 50 measurement shots, respectively. For each random seed, 200 test Hamiltonians are sampled, and results are aggregated over 10 seeds.}
\label{fig:2D SKQD result}
\end{figure*}

To demonstrate the practical significance of EGATE-NNVQE, we evaluate its utility as a state-preparation mechanism for quantum eigensolvers with stronger theoretical guarantees. While VQE is widely used as a variational heuristic, its convergence properties are not rigorously characterized in general. In contrast, QSMs and QPE offer systematic convergence guarantees, provided that the initial reference state has sufficient overlap with the true ground state. From this perspective, the quality of the initial state becomes the critical factor determining downstream performance. We therefore investigate whether the states generated by EGATE-NNVQE serve as more effective reference states for SKQD, a representative QSM, compared to the NNVQE baseline or random initialization.


We briefly review QSMs and refer the reader to Refs.~\cite{KQD_1, KQD_2, kim2023evidence,QSCI,SQD,SKQD} for detailed treatments. At a high level, QSMs construct a low-dimensional subspace, project the Hamiltonian 
$H$ onto this subspace, and classically diagonalize the projected Hamiltonian. Different QSM variants primarily differ in how the subspace is constructed.

In KQD~\cite{KQD_1,KQD_2}, the corresponding subspace is the unitary Krylov space $\mathcal{K}_U^d$
\begin{equation}
\mathcal{K}_U^d = \text{Span}\lbrace |\psi_k\rangle\rbrace_{k=0}^{d-1},\text{ where } |\psi_k\rangle =e^{-iH(k dt)} |\psi_0\rangle,
\end{equation}
where $|\psi_0\rangle$ is the initial (reference) state. Then the Hamiltonian is projected onto $\mathcal{K}_U^d$, and low-energy eigenvalues are obtained by solving a $d\times d$ generalized eigenvalue problem for the projected Hamiltonian, where $d\ll 2^n$.
In SQD, the subspace $\mathcal{S}$ is constructed via quantum sampling, i.e., quantum-selected configuration interactions~\cite{QSCI,SQD}. Specifically, a quantum circuit prepares a reference state $|\psi_0\rangle$, and $\mathcal{S}$ is defined as
\begin{align}
&\mathcal{S} = \text{Span}\lbrace |x\rangle : x\in\mathcal{X}\rbrace,\notag \\
&\text{ where }\mathcal{X} = \lbrace x|\mathrm{P}_{\psi_0}(x)= \lvert \langle x|\psi_0 \rangle\rvert^2 \rbrace.
\end{align}
The Hamiltonian is then projected onto $\mathcal{S}$, and the ground-state energy is estimated by solving the resulting low-dimensional eigenvalue problem.

SKQD~\cite{SKQD}  combines ideas from KQD and SQD by sampling from the Krylov basis, thereby leveraging the strengths of both approaches. Given a Krylov dimension $d$, time step $dt$ and an initial state $|\psi_0\rangle$, Krylov basis states are constructed as $\lbrace|\psi_k\rangle := e^{-iH(k dt)}|\psi_0\rangle\rbrace_{k=0}^{d-1}$. For each $k\in\lbrace0,1,\dots,d-1\rbrace$, $M$ computational-basis measurements of $|\psi_k\rangle$ are performed, yielding a sequence of bitstrings $\lbrace |a_{km}\rangle\rbrace_{m=0}^{M-1}$, where $a_{km}\sim P_{\psi_k}(x)=|\langle x|\psi_k\rangle|^2$.  
The sampled subspace is then defined as
\begin{equation}
\mathcal{S}_{d,M} :=\text{Span} \lbrace|a_{km}\rangle | k=0,\dots,d-1; m=0,\dots,M-1 \rbrace.
\end{equation}
Finally, the ground-state energy is estimated by solving the low-dimensional eigenvalue problem for the Hamiltonian projected onto $\mathcal{S}_{d,M}$.

Under appropriate conditions, QSMs can provide ground-state energy estimates of the form $\Pr[|\tilde{E}_0-E_0|<\epsilon]\ge 1-\delta$~\cite{KQD_1,KQD_2,SKQD}, where $\tilde{E}_0$ ($E_0$) denotes the estimated (true) ground-state energy and $1-\delta$ is the confidence for target precision $\epsilon$.
Krylov-based methods typically require that the initial state has at least inverse-polynomial overlap with the true ground state, as well as a reasonably well-conditioned low-energy spectrum (e.g., not an extremely small gap). 
For sample-based variants, performance depends not only on the quality of the initial state but also on the measurement-basis sparsity of the true ground state, i.e., whether most of its probability mass is concentrated on a relatively small (polynomially bounded) number of bitstrings. 
Accordingly, SKQD is most effective when the initial state has inverse-polynomial overlap with the true ground state and the true ground state satisfies the sparsity condition~\cite{SKQD}.
In practice, once sparsity conditions are satisfied, the critical factor determining performance is the overlap (or fidelity) between the initial state and the true ground state. 

We evaluate the effectiveness of EGATE-NNVQE as an initializer for SKQD by benchmarking three input-state strategies---Haar-random state, the NNVQE baseline, and EGATE-NNVQE---across four Hamiltonian families, Eqs.~(\ref{eq:XXZ}), (\ref{eq:XXZ+X}), (\ref{eq:2D_XXZ}), and (\ref{eq:2D_XYZ}).
We set the maximum Krylov dimension to $d=10$ and use a uniform time step $dt=\pi/\|H\|$, where $\|\cdot\|$ denotes the operator norm. The Krylov basis states $|\psi_k\rangle = e^{-iH k dt}|\psi_0\rangle$ for $k\in\{0,\ldots,d-1\}$ are implemented via first-order Trotterization with 10 steps. For each $k$, we perform $M\in\{25,50\}$ computational-basis measurements to sample basis states and construct the SKQD subspace; thus, $M$ denotes the number of shots per Krylov dimension $k$.
All SKQD experiments are conducted under a realistic noise model constructed from the IBM Quantum \texttt{FakeSherbrooke} backend.

Figures~\ref{fig:1D SKQD result} and~\ref{fig:2D SKQD result} report SKQD results for two 1D Hamiltonian families, $H_{\mathrm{XXZ}}$ and $H_{\mathrm{XXZ+X}}$, on $n=8$ qubits, and for two 2D families on a $3\times3$ lattice, $H_{\mathrm{XXZ}}^{3\times3}$ and $H_{\mathrm{XYZ}}^{3\times3}$, respectively. The left and right columns use $M=25$ and $M=50$, respectively.
Each curve shows the mean $\pm$ standard deviation (SD) over 10 random seeds. For each seed, we randomly sample a set of test Hamiltonian instances: in Fig.~\ref{fig:1D SKQD result}, we sample 100 instances for $H_{\mathrm{XXZ}}$ and 200 for $H_{\mathrm{XXZ+X}}$, while in Fig.~\ref{fig:2D SKQD result}, we sample 200 instances per seed for each 2D Hamiltonian family.
For the random baseline (blue), each seed additionally corresponds to an independently sampled Haar-random state used directly as the SKQD initial state. 
For NNVQE (orange) and EGATE-NNVQE (green), the initial states are obtained by inference using the pretrained model trained with the same seed in Sec.~\ref{sec:generalization}, without any instance-specific optimization.

The two-local Hamiltonians considered in our SKQD benchmark empirically exhibit sufficient effective sparsity for SKQD to be applicable. Consequently, the dominant factor in practice is the quality (fidelity) of the initial state: better initializers achieve smaller relative energy errors with fewer shots and Krylov dimensions for a fixed accuracy target. As shown in Figs.~\ref{fig:1D SKQD result} and~\ref{fig:2D SKQD result}, EGATE-NNVQE consistently reduces the shot requirement and the Krylov dimension needed to reach a target error compared to the random and NNVQE  baselines, indicating that its inferred states provide higher-quality initialization for SKQD.

While our demonstration focuses on near-term QSMs such as SKQD, the same high-overlap initializer is also well suited for QPE in a fault-tolerant setting. In QPE,
the success probability $P=\lvert\langle\psi_{0}|\psi_{\mathrm{ground}}\rangle\rvert^2$ directly determines the repetition overhead. A larger overlap therefore reduces the number of repetitions $M \ge -\ln(\frac{1}{\delta})/\ln(1-P)\approx\ln(\frac{1}{\delta})/P$ required to achieve confidence $1-\delta$.
Since EGATE-NNVQE generalizes across a Hamiltonian family, a single trained model can initialize QPE for many unseen instances without additional instance-specific optimization, improving end-to-end resource efficiency.


\begin{figure*}
\centering
\begin{tabular}{c c}
    \includegraphics[width=0.46\textwidth]{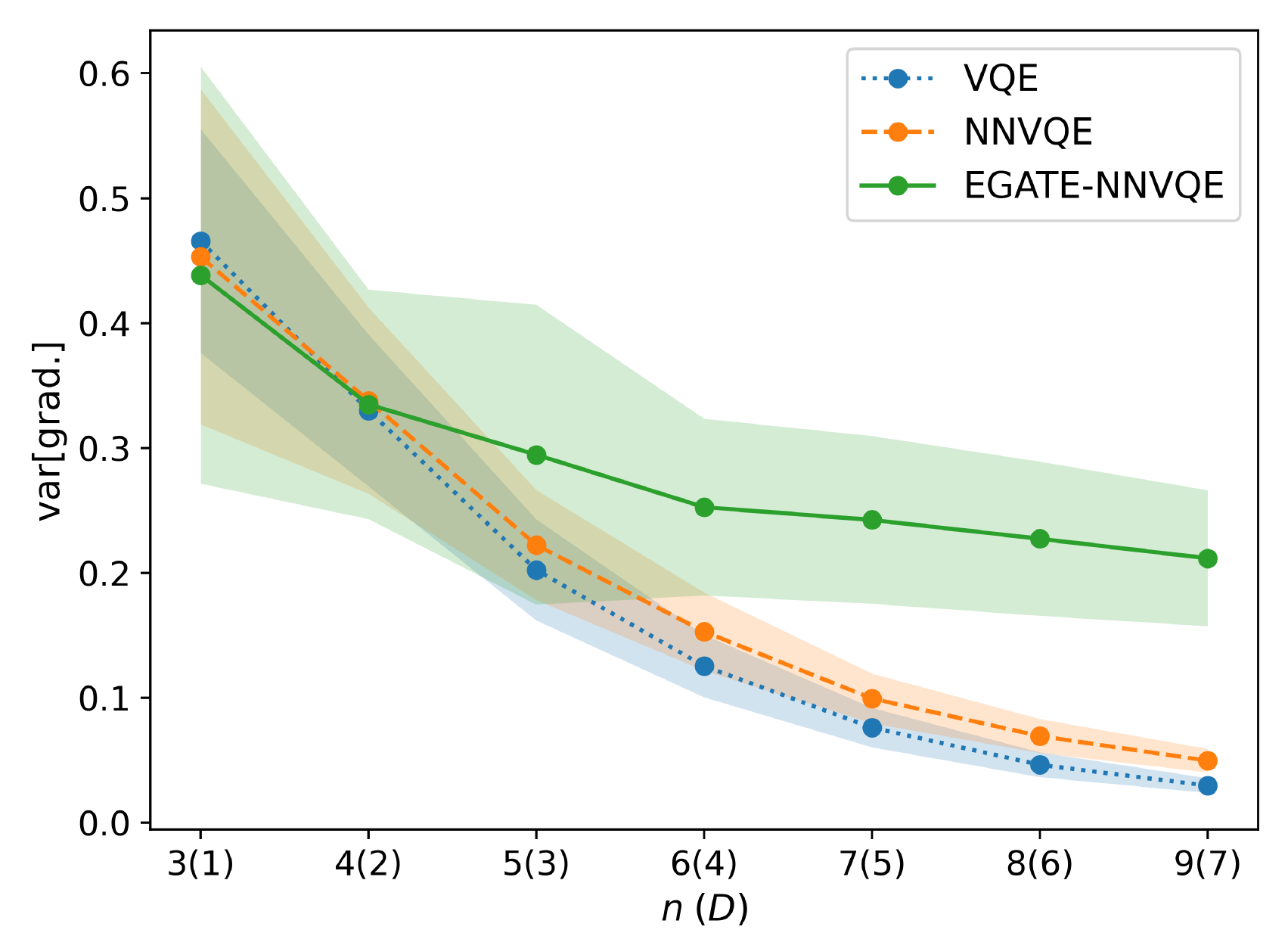} &
    \includegraphics[width=0.46\textwidth]{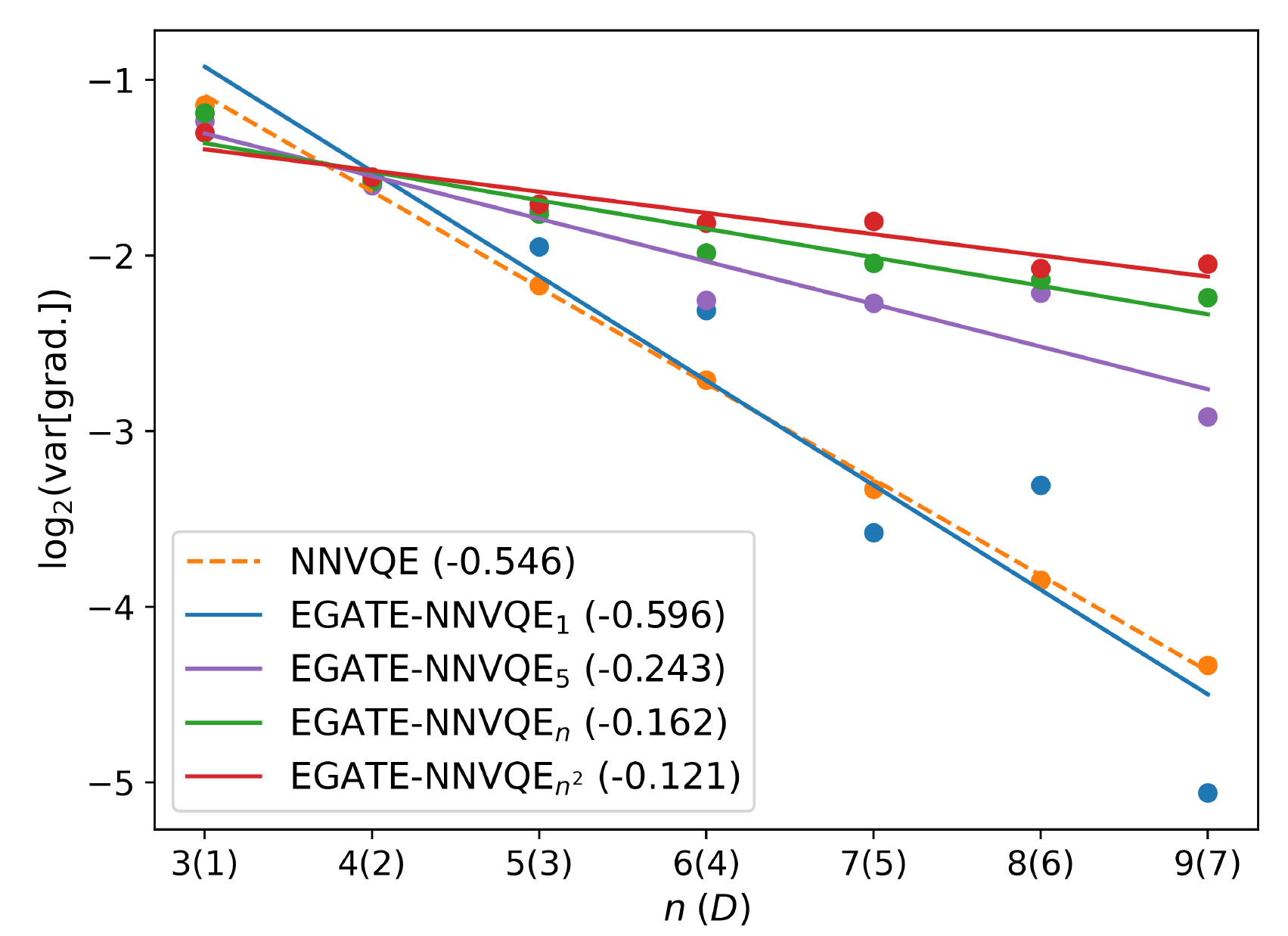} \\
    (a) & (b)
  \end{tabular}
  \caption{Comparison of BP behavior across VQE variants for the 1D XXZ Heisenberg spin chain (Eq.~(\ref{eq:XXZ})) with $J^{zz} = 1$. (a) Gradient variance, $\text{var}[\text{grad.}]$, as a function of the number of qubits $n$ (with corresponding ansatz block depth $D$) for three approaches: standard VQE (blue), NNVQE (orange), and EGATE-NNVQE (green). While NNVQE provides minimal mitigation of the BP problem, incorporating EGATE significantly improves robustness by preserving larger gradient variance at increasing problem sizes. (b) Log-scaled gradient variance, $\log_2( \text{var}[\text{grad.}])$, together with linear fits for EGATE-NNVQE with different latent-space dimensionalities. Subscripts in the legend indicate the latent dimension, and values in parentheses denote the fitted slopes.}
\label{fig:BP result}
\end{figure*}

\subsection{Barren Plateaus}
\label{sedc:BP}
The BP phenomenon leads to an increasingly flat optimization landscape and is a significant obstacle to practical VQE training~\cite{mcclean2018barren, wang2021noise, larocca2025barren}. To diagnose BP, we examine how variance in parameter gradients decays as the system size $n (D)$ increases and systematically compare the scaling behavior across standard VQE, NNVQE baseline, and EGATE-NNVQE.  
To isolate BP effects from hardware noise and hyperparameter choices, we run all experiments on the IBM noiseless simulator, keeping the ansatz and network architectures identical across baselines.

For each $n (D)$, we fix a single Hamiltonian instance and perform 500 independent trials that differ only in random initialization (i.e., ansatz parameter initialization for VQE and NN weight initialization for NNVQE and EGATE-NNVQE; input of NN is the same for each configuration). In each trial, we compute the gradient of all trainable parameters at the first optimization step.
For each ansatz parameter, we calculate the variance of its gradient across the 500 trials. This results in a variance vector of dimension equal to the number of ansatz parameters, where each element corresponds to the variance of a specific parameter’s gradient. The plotted marker is the mean of this variance vector, i.e., the average variance across all ansatz parameters. The shaded region represents the standard deviation of the variance vector, capturing the variability in how severely each parameter is affected by BPs.

Unlike the generalization experiments, EGATE is trained once for each $n (D)$ on the fixed instance to obtain a single latent vector, which is held fixed across the 500 trials. In this setting, EGATE serves purely as a data-shifting module that maps the fixed Hamiltonian structure to a fixed latent representation, while the only source of randomness across trials is the parameter initialization. This controlled setup is therefore well-suited for isolating the role of EGATE in mitigating BPs, as BP diagnostics target gradient statistics under random initialization rather than test-time generalization. By fixing both the Hamiltonian and the EGATE conditioning, we remove instance-level and encoder-training variability, enabling a controlled comparison of gradient-variance scaling with $n$.

Figure~\ref{fig:BP result}(a) reports the BP behaviour of standard VQE (blue), NNVQE (orange), and EGATE-NNVQE (green) as $n$ (and $D$) increases. Here, we consider $H_\mathrm{XXZ}$, Eq.~\ref{eq:XXZ}), with $J^{zz} = 1$ as a target Hamiltonian, and NNVQE is implemented with one hidden layer.
For EGATE, we adopt the same H-graph construction as in Fig.~\ref{fig:1D_generalization_result}(a): one-hot node features and edge features $(1,1,J^{zz})$. With this configuration, EGATE produces a latent vector of dimension $n+3$. 
As $n (D)$ grows, all methods exhibit a clear decline in gradient variance. Notably, the NNVQE baseline shows BP behavior comparable to standard VQE, whereas EGATE-NNVQE exhibits a markedly slower decay, indicating improved robustness to BPs. 

The key difference among the three algorithms lies in the ansatz parameter initialization mechanism: standard VQE samples parameters uniformly from $[0,2\pi]$, while NNVQE and EGATE-NNVQE generate parameters using NN with weights initialized from a standard normal distribution $\mathcal{N}(0,1)$. However, the two models differ in their inputs: NNVQE takes a simple parameter vector $J^{zz}$ as input, whereas EGATE-NNVQE processes a H-graph encoding richer structural information. In other words, the structure of the EGATE-NNVQE input introduces an inductive bias that influences how the NN generates the ansatz initialization. 
Furthermore, as demonstrated in Sec.\ref{sec:generalization}, EGATE can extract structural information from Hamiltonians that become increasingly complex as the $n (D)$ grows and correspondingly, our BP results show a widening performance gap with $n (D)$.
As a result, even when the NN architecture remains fixed, EGATE-NNVQE produces a more informative initialization of the ansatz parameters than the NNVQE baseline. In the VQA literature, such improved initialization is often referred to as a ``good initialization” or ``warm start”, which is known to alleviate the effects of BPs and stabilize training~\cite{larocca2025barren}.

To further substantiate this interpretation, we perform linear fits on a $\text{log}_2$ scale to the variance of the gradient for NNVQE and EGATE-NNVQE under different latent-dimensionality settings (subscripts denote latent dimension), in Fig.~\ref{fig:BP result}(b). NNVQE and EGATE-NNVQE$_n$ share the same setting as in plot (a). For EGATE-NNVQE$_1$ and EGATE-NNVQE$_5$, we obtain constant-dimensional latents via a linear projection after sum pooling, whereas EGATE-NNVQE$_{n^2}$ uses concatenation merge, Eq.~(\ref{eq:concat merge layer}), and MLP pooling. EGATE-NNVQE$_1$ exhibits similar BPs to those of the NNVQE baseline, while larger and more expressive latents reduce the fitted slope, indicating a slower decay of gradient variance. These results suggest that informative Hamiltonian representations are crucial for BP mitigation.

\section{Conclusion and Future Work}


In this work, we introduced a general protocol for representing arbitrary one- and two-local Pauli Hamiltonians as undirected H-graphs with multi-dimensional node and edge features. Building on this representation, we proposed EGATE, an edge-featured graph attention autoencoder that learns structure-aware embeddings of H-graphs in an unsupervised manner. 
To isolate the effect of representation learning, our benchmarks focused on controlled interaction topologies. Nevertheless, the proposed H-graph construction and EGATE are Hamiltonian-agnostic and apply to any qubit Hamiltonian with at most two-local terms. Because EGATE is trained without supervision, it requires no labeled training data (e.g., VQE-optimized ansatz parameter targets~\cite{zhang2025qraclegraphneuralnetworkbasedparameterinitializer}), eliminating the costly resource overhead for training data generation entirely.  

Experimental results demonstrate that EGATE-NNVQE substantially improves generalization and enhances trainability by exhibiting greater robustness to BP phenomenon. 
Importantly, EGATE-NNVQE is not intended to function as a standalone eigensolver capable of producing accurate ground states across an entire Hamiltonian family without adaptation.
Rather, it is best viewed as a high-quality initializer that generates initial states with enhanced ground-state overlap for previously unseen instances. This role naturally complements downstream eigenvalue-estimation routines such as QPE and QSMs, whose efficiency depends critically on the initial overlap. We validated this role by warm-starting SKQD in Sec.~\ref{Sec:SKQD}, where EGATE-NNVQE initial states yield improved convergence under noisy simulation. 
These results position EGATE-NNVQE as a practical tool for both near-term quantum devices and fault-tolerant architectures.

From an architectural standpoint, EGATE provides a flexible framework that can be extended with alternative components and training strategies.
Future work may explore more expressive merge and pooling mechanisms, such as sequential modules to capture layer-wise structure, attention-based pooling or graph explicit pooling~\cite{liu2023graphpoolinggraphneural,li2024graph,liu2025graph} to reflect node and edge importance, and alternative decoders from the GAE literature~\cite{kipf2016variationalgraphautoencoders,park2019symmetric} to further improve representational capacity. 
Another promising direction is to couple graph representation learning with downstream objectives, such as ground-state energy estimation, through joint or guided training strategies (e.g., guided graph compression~\cite{casals2025guidedgraphcompressionquantum}), thereby directly optimizing latent representations for task performance.
Although the current H-graph representation requires at most quadratic classical memory in the number of qubits 
$n$ and is therefore typically manageable, extremely large systems may motivate exploring quantum analogues of EGATE based on quantum graph neural networks~\cite{QGNN,QGAT}. Such approaches could provide complementary scalability advantages in regimes where classical preprocessing becomes limiting.

From a deployment perspective, using SKQD and related routines on real devices favors shallow ans\"{a}tze because noise accumulates rapidly with depth. 
At the same time, improved generalization often benefits from more expressive (typically deeper) ans\"{a}tze.
Importantly, even if the classical neural predictor is made substantially more expressive, its generalization capability remains fundamentally constrained by the representational capacity and trainability of the underlying ansatz. This suggests limited gains from increasing the complexity of the classical network alone. A practical direction is therefore to optimize the ansatz circuit~\cite{grimsley2019adaptive, ostaszewski2021reinforcement, du2022quantum, dai2024quantum} under fixed depth budgets while accounting for the target Hamiltonian family and hardware constraints, such as connectivity and native gate sets.

More broadly, this work highlights the potential of structure-aware representation learning as a foundational component for scalable quantum algorithms.

\section*{Acknowledgments}
This work is supported by Institute of Information \& communications Technology Planning \& evaluation (IITP) grant funded by the Korea government (No. 2019-0-00003, Research and Development of Core Technologies for Programming, Running, Implementing and Validating of Fault-Tolerant Quantum Computing System), the National Research Foundation of Korea (RS-2025-02309510), the Ministry of Trade, Industry, and Energy (MOTIE), Korea, under the Industrial Innovation Infrastructure Development Project (RS-2024-00466693), and by Korean ARPA-H Project through the Korea Health Industry Development Institute (KHIDI), funded by the Ministry of Health \& Welfare, Korea (RS-2025-25456722).

\newpage
\bibliographystyle{ieeetr}
\bibliography{reference}

\appendix
\setcounter{section}{0}
\setcounter{figure}{0}
\setcounter{table}{0}
\setcounter{equation}{0}
\setcounter{algorithm}{0}

\renewcommand{\thesection}{\Alph{section}} 
\renewcommand\thefigure{\Alph{section}.\arabic{figure}} 
\renewcommand\theequation{\Alph{section}.\arabic{equation}} 
\renewcommand\thetable{\Alph{section}.\arabic{table}} 
\renewcommand\thealgorithm{\Alph{section}.\arabic{algorithm}}

\twocolumn[{
  \centering
  \LARGE\bfseries Appendix\par
  \vspace{0.8em}
}]

\section{Algorithmic Architecture}
\label{appendix:sec:architecture}
\begin{algorithm}[t]
\caption{Node Module}
\label{appendix:alg:node module}
\textbf{Input}: Node features $\textbf{O}^{l-1}\in\mathbb{R}^{n\times (d_o+d_s)}$, edge features $\textbf{E}^{l-1}\in\mathbb{R}^{m\times d_e}$, contribution ratio $\lambda$\\
\textbf{Parameter}: learnable parameters $\textbf{W}^{o}\in\mathbb{R}^{\lfloor \lambda d_o \rfloor \times d_o}, \textbf{W}^{e}\in\mathbb{R}^{\lceil (1-\lambda) d_o \rceil \times d_e}$, and edge-integrated attention mechanism $\mathbf{a}: \mathbb{R}^{2\lfloor \lambda d_o \rfloor +\lceil (1-\lambda) d_o \rceil}\mapsto\mathbb{R}
$\\
\textbf{Output}: Updated node features $\textbf{O}^l\in\mathbb{R}^{n\times (d_o+d_s)}$

\begin{algorithmic}[1]
\STATE $\textbf{O}^{*} \leftarrow \textbf{W}^{o}\textbf{O}^{l-1}$; $\textbf{E}^{*} \leftarrow W^{e}\textbf{E}^{l-1}$ 
\STATE Let $\mathcal{N}_i$ represents the neighboring nodes of node $i$, and $\sigma$ and $\sigma'$ are non-linearity functions, e.g. LeakyReLU, ELU.
\FOR{each node $i$}
    \FOR{each $j \in \mathcal{N}_i$}
        \STATE $w_{ij} \leftarrow \sigma\left(\mathbf{a}^T\left[\vec{o}^*_i \mathbin\Vert \vec{o}^*_j \mathbin\Vert \vec{e}^{*}_{ij}\right]\right)$
    \ENDFOR
    \STATE $\alpha_{ij} \leftarrow \mathrm{softmax}_{j}(w_{ij})$
    \STATE $\vec{o}^l_i \leftarrow \sigma' \left(\sum_{j \in \mathcal{N}_i} \alpha_{ij}\left[\vec{o}^*_j \mathbin\Vert \vec{e}^{*}_{ij}\right]\right)$
\ENDFOR
\STATE \textbf{return} $\textbf{O}^l$
\end{algorithmic}
\end{algorithm}

\subsection{EGATE}
\label{appendix:sec:EGATE}

\paragraph{Node Module \& Edge Module}
The primary objective of our proposed edge-featured graph attention autoencoder (EGATE) in the main paper is to learn a compact and informative representation $\vec{g}$ of the Hamiltonian-graph $\textbf{ G}=(\textbf{O},\textbf{E})$. 
The edge-featured graph attention (EGAT) layer~\cite{EGAT} serves as a core component of EGATE, as described in Fig.~\ref{fig:model} of the main manuscript. The EGAT layer simultaneously captures and aggregates both node- and edge-level information via two modules: the node module and the edge module. We give pseudocode for each module in the algs.~\ref{appendix:alg:node module} and \ref{appendix:alg:edge module}, respectively. 

The $l$th EGAT layer takes the representations produced by its predecessor layer, $(\textbf{O}^{l-1}, \textbf{E}^{l-1})$, and generates refined descriptors, $(\textbf{O}^l, \textbf{E}^l)$. Specifically, the node module for $l$th EGAT layer takes $(\textbf{O}^{l-1}, \textbf{E}^{l-1})$ as input and updates node representations as $\textbf{O}^l$ using an edge-integrated attention mechanism $\mathbf{a}$ that incorporates edge features into both message passing and attention score computation. 
The edge module mirrors the same attention scheme under a node-transit strategy, followed by an MLP that transforms the aggregated features into higher-level edge representations. Note that edge module takes $(\textbf{O}^{l}, \textbf{E}^{l-1})$ as input and updates edge representation as $\textbf{E}^{l}$. 
The contribution ratio $\lambda$ controls the relative weighting of node versus edge features in the message passing process.
Throughout all experiments, we fix $\lambda = 0.5$ to weight the two equally; choosing a smaller (larger) $\lambda$ increases the relative contribution of edge (node) features.

The total number of EGAT layers \$L\$ is a hyperparameter, and we empirically observe that the EGATE test loss saturates beyond a moderate number of layers and additional layers provide only marginal improvement; a similar trend that has also been reported in the single GNN settings~\cite{EGAT}.
In our work, we adopt an MLP with two hidden layers as the edge-integrated attention mechanism $\mathbf{a}$ and edge update network in step 10 of Alg.~\ref{appendix:alg:node module}.


\begin{algorithm}[t]
\caption{EdgeModule}
\label{appendix:alg:edge module}
\textbf{Input}: Node features $\textbf{O}^{l}\in\mathbb{R}^{n\times (d_o+d_s)}$, edge features $\textbf{E}^{l-1}\in\mathbb{R}^{m\times d_e}$, contribution ratio $\lambda$ \\
\textbf{Parameter}: learnable parameters $\textbf{W}^{o}\in\mathbb{R}^{\lfloor \lambda d_o \rfloor \times d_o}, \textbf{W}^{e}\in\mathbb{R}^{\lceil (1-\lambda) d_o \rceil \times d_e}$, and edge-integrated attention mechanism $\mathbf{a}: \mathbb{R}^{2\lfloor \lambda d_o \rfloor +\lceil (1-\lambda) d_o \rceil}\mapsto\mathbb{R}
$,  which are not shared among node modules\\
\textbf{Output}: Updated edge features $\textbf{E}^l\in\mathbb{R}^{m\times d_e}$

\begin{algorithmic}[1]
\STATE Let, $\sigma$ indicates LeakyReLU
\STATE $\textbf{O}^{*} \leftarrow \textbf{W}^{o}\textbf{O}^l$;\quad $\textbf{E}^{*} \leftarrow \textbf{W}^{e}\textbf{E}^{l-1}$

\FOR{each node $i$}
    \FOR{each $j\in\mathcal{N}_i$}
        \STATE $\beta_{ij} \leftarrow \mathrm{softmax}_{j}\left(\sigma\left(\mathbf{a}^T\left[\vec{o}^*_i \mathbin\Vert \vec{o}^*_j \mathbin\Vert \vec{e}^{*}_{ij}\right]\right)\right)$
    \ENDFOR
   \STATE $\vec{e}'_{i} \leftarrow \sum_{j\in\mathcal{N}_i} \beta_{ij}\vec{e}^{*}_{ij} $ $\gets$ Node-transit strategy
    \ENDFOR

\FOR{each edge $(i,j)$}
    \STATE $\vec{e}^l_{ij} \leftarrow \mathrm{MLP}\left( \vec{o}^*_i, \vec{o}^*_j, \vec{e}'_i, \vec{e}'_j, \vec{e}^{l-1}_{ij}
    \right)$
\ENDFOR

\STATE \textbf{return} $\textbf{E}^l$
\end{algorithmic}
\end{algorithm}

\paragraph{Decoder} The decoder of EGATE is implemented as an MLP with one hidden layer, intentionally restricting its excessive expressivity and thereby encouraging the encoder to learn informative embeddings. The number of hidden units is the hyperparameter.


\subsection{Ansatz Structure}
\label{appendix:sec:ansatz}
\begin{figure*}[tb]
\centering
\begin{tabular}{c}
    \includegraphics[width=0.9\textwidth]{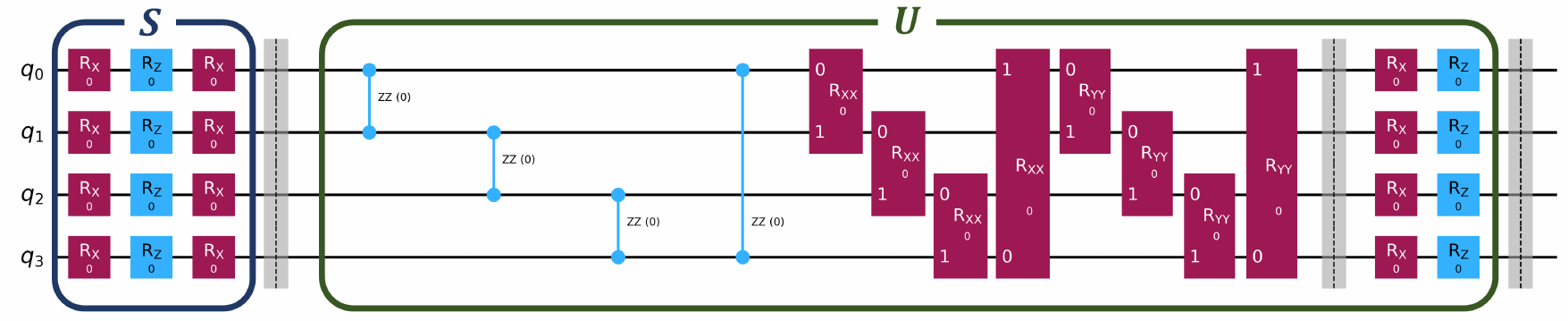} \\
    (a)\\
    
    \includegraphics[width=0.9\textwidth]{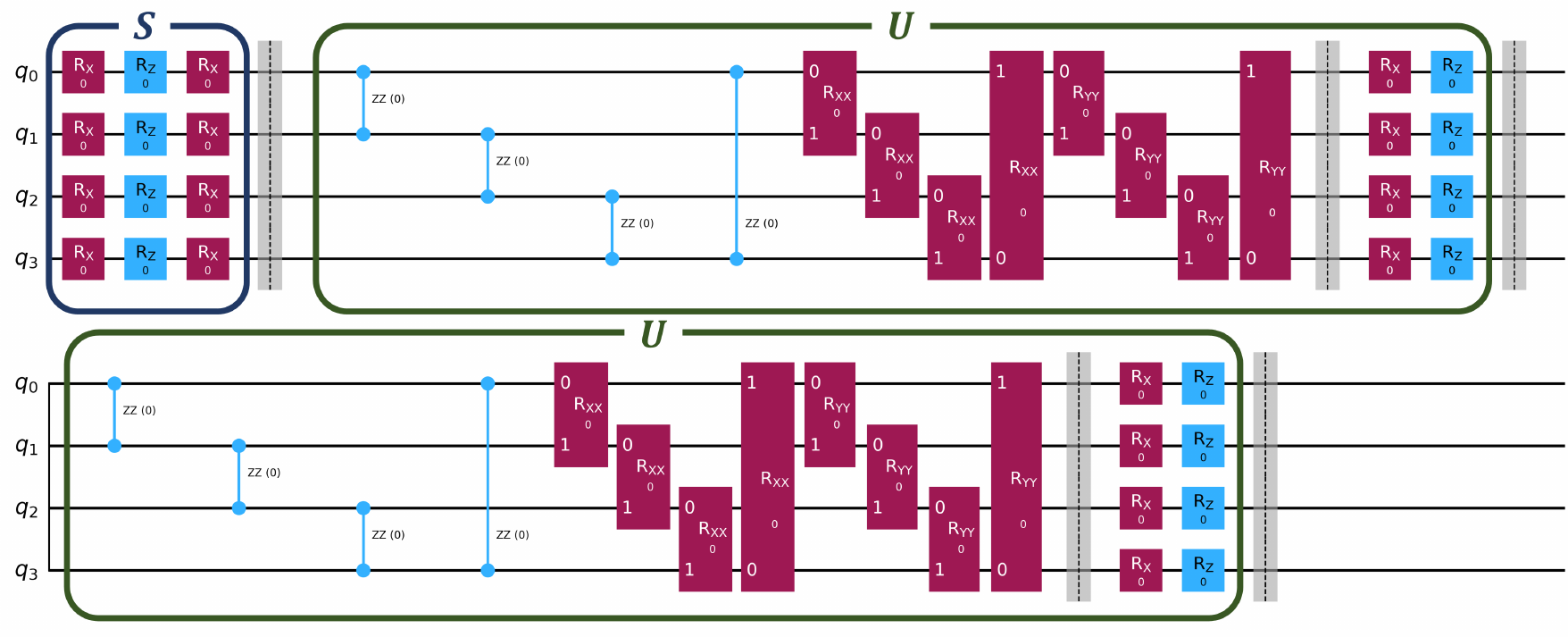} \\
    (b)\\    
  \end{tabular}
\caption{4-qubits ladder-wise hardware-efficient ansatz (HEA) structure: (a) shows $D=1$ HEA, (b) depicts $D=2$ HEA.
The blue box denoted by $S$ is a universal single-qubit rotation operating on all qubits, whereas the green box indicated by $U$ is
a unitary block for the ansatz. The number of repetitions of $U$ in the ansatz is denoted by $D$.}
\label{appendix:fig:ansatz}
\end{figure*}

The ansatz circuit employed in this work is the ladder-wise hardware-efficient ansatz (HEA)~\cite{HEA}, which is the same one used in the original neural-network-encoded variational quantum eigensolver (NNVQE) paper~\cite{NN-VQA}. 
The state after ladder-wise HEA can be formulated as, 
\begin{equation}
    |\psi_D\rangle = U|\psi_{D-1}\rangle = \cdots = U^{D}|\psi_0\rangle =  U^{ D} \left(S\,|0\rangle^{\otimes n}\right),
\end{equation}
where $n$ denotes the number of qubits, $D$ is the number of repetitions of the ansatz block $U$, and $S  = \bigotimes_{i=1}^n R^{(i)}_x(\theta^x_{i2}) R^{(i)}_z(\theta^z_{i}) R^{(i)}_x(\theta^{x}_{i1})$ indicates a universal single-qubit rotation operating on all qubits as depicted in Fig.~\ref{appendix:fig:ansatz} with the blue box. The superscript $i$ indicates that the single-qubit rotation acts on the qubit $i$.
The ansatz block of HEA is 
\begin{align}
    U = &\left[\bigotimes_{i=1}^n R^{(i)}_z({\theta}_i^z) R^{(i)}_x({\theta}_i^x) \right] 
    \times \notag\\
    &\left[\begin{aligned}
        \prod _{i=1}^n R^{(i,i+1)}_{yy}({\theta}_{i, i+1}^{yy})
            \prod _{i=1}^n R^{(i,i+1)}_{xx}({\theta}_{i, i+1}^{xx})\\
            \prod _{i=1}^n R^{(i,i+1)}_{zz}({\theta}_{i, i+1}^{zz})  
    \end{aligned}
    \right],
    \label{eq:unitary_block}
\end{align}
where $R^{(i)}_\alpha$ and $R^{(i,i+1)}_{\alpha \alpha}$ is the single- and two-qubits rotation gates along $\alpha$ direction for $\alpha\in\lbrace x,y,z\rbrace$. The superscript $(i,i+1)$ specifies that the two-qubit rotation acts on the adjacent qubits $i$ and $i+1$; site indices are taken modulo $n$, i.e., $(i=n)+1 = 1$. 
Parameterized two-qubit gates, $R_{xx}$, $R_{yy}$, and $R_{zz}$, are arranged in a ladder-wise pattern, or nearest-neighbor loop pattern, as illustrated in the 4-qubits example HEA in Fig.~\ref{appendix:fig:ansatz}.

\subsection{NN structure for NNVQE}
\label{appendix:sec:NN structure}
A fully connected MLP with one or two hidden layers is used as a neural network (NN) in NNVQE. All hidden layers use ReLU activations, while the output layer uses a sigmoid activation; the resulting outputs are then scaled by $2\pi$ to yield ansatz parameters in $[0,2\pi]$.
The input dimension is set to the number of tunable Hamiltonian parameters for the NNVQE baseline, and to the dimension of the graph latent vector $\vec{g}$  for EGATE-NNVQE and input-expanded NNVQE. In input-expanded-NNVQE, the tunable parameter vector is replicated to match the dimension of $\vec{g}$. 
The output dimension equals the number of ansatz parameters.
The number of hidden units is a hyperparameter that originally varies with the ansatz block depth $D$~\cite{NN-VQA}. In our experiments, we mostly use the network width of 20 units for the one hidden layer architecture and 20$\rightarrow$40 units for the two hidden layer architecture, ensuring consistent capacity and enabling fair comparisons across all systems. For practical use, however, this hyperparameter could be tuned appropriately (and we recommend it).
Unless stated otherwise, ``NN” collectively refers to the networks used in NNVQE and in the NNVQE component of EGATE-NNVQE and input-expanded-NNVQE; these methods share the same hidden layer architecture and differ only in the input.


\section{Experimental Setting}
\label{appendix:sec:experimental setting}
\subsection{Generalization}
In this section, we describe the specific hyperparameter settings used for comparing generalization performance on four different Hamiltonian families, $H_{\mathrm{XXZ}},H_{\mathrm{XXZ+X}}$, $H_{\mathrm{XXZ}}^{3\times3}$, and $H_{\mathrm{XYZ}}^{3\times3}$, respectively.


\paragraph{$H_{\mathrm{XXZ}}$}: 
We simulate 4-, 6-, and 8-qubit instances of $H_{\mathrm{XXZ}}$ (Eq.~(\ref{eq:XXZ})) using NNVQE with one or two hidden layers. We additionally include input-expanded-NNVQE to verify that merely increasing the input dimensionality does not improve generalization performance. For NNVQE, EGATE-NNVQE, and input-expanded-NNVQE, the training set for the tunable coupling $J^{zz}$ consists of 20 equally spaced points in $[-3,3]$, and performance is evaluated on 1,000 equally spaced test points in $[-10,10]$ for each system size. 

Each Hamiltonian is encoded as an H-graph with one-hot node features $\vec{o}_i=\mathbf{e}_i\in\mathbb{R}^n$ and edge features $(1,1,J^{zz})$.  EGATE uses 5, 7, and 9 EGAT layers for $n=4,6,8$, respectively, and its decoder is a one hidden layer MLP with 18, 32, and 55 hidden units. EGATE is trained on the same training set for 100 epochs with a learning rate of $0.001$, without mini-batch training or a scheduler.

NNVQE baseline takes $J^{zz}$ as input, whereas EGATE-NNVQE uses a graph-level embedding $\vec{g}$ of dimension 7, 9, and 11 for $n=4,6,8$, respectively. 
For each $n$, we use an $n$-qubit ladder-wise HEA with depth $D=2$, resulting in 52, 78, and 104 ansatz parameters for $n=4,6,8$, respectively. 
Throughout all generalization experiments, the NN weights are initialized from a normal distribution with standard deviation 0.1 as $\mathcal{N}(0.0,0.1)$.
NNVQE is trained with a learning rate of 0.003 for up to 200 iterations, without mini-batch training or a learning rate scheduler.



\paragraph{$H_{\mathrm{XXZ+X}}$}: 
We simulate 4-, 6-, and 8-qubit instances of $H_{\mathrm{XXZ+X}}$ (Eq.~(\ref{eq:XXZ+X})) using NNVQE with one or two hidden layers.  For both NNVQE and EGATE-NNVQE, the training set for the tunable couplings $J^{zz}$ and $K^x$ consists of 20 equally spaced points in $[-3,3]$ per parameter, and performance is evaluated on 200 equally spaced test points in $[-10,10]$ per parameter, yielding 400 training ans 40,000 test Hamiltonian for each system size.

Hamiltonian is encoded as an H-graph with permutation-invariant node features $\vec{o}_i=(1,K^x)$ and edge features $\vec{e}_{ij}=(1,1,J^{zz})$.
EGATE uses 5, 7, and 9 EGAT layers for $n=4,6,8$, respectively, and its decoder is a one hidden layer MLP with 18, 32, and 55 hidden units.
EGATE is trained using mini-batch training with 10 batches, where the starting learning rate is 0.001, and it decays to its 80\% every 10 steps. The maximum iteration is 50.

NNVQE takes $(J^{zz},K^x)$ as input, whereas EGATE-NNVQE uses a graph-level embedding $\vec{g}$ with a fixed 5-dimensional latent vector independent of $n$. For each $n$, we use an $n$-qubit ladder-wise HEA with depth $D=2$, resulting in 52, 78, and 104 ansatz parameters for $n=4,6,8$, respectively. NNVQE is trained with a learning rate of 0.003 for up to 200 iterations, without mini-batch training or a learning rate scheduler.




\paragraph{$H_{\mathrm{XXZ}}^{3\times3}$}: 
We simulate the 9-qubit Hamiltonian $H_{\mathrm{XXZ}}^{3\times 3}$ (Eq.~(\ref{eq:2D_XXZ})), defined on a $3\times 3$ (2D) lattice, using NNVQE with two hidden layers.
For both NNVQE and EGATE-NNVQE, the training set for the tunable couplings $J^{zz}$ and $K^x$ consists of 20 equally spaced points in $[-3,3]$ per parameter, and performance is evaluated  on 200 equally spaced test points in $[-10,10]$ per parameter, yielding 400 training and 40,000 test Hamiltonians.

Hamiltonian is encoded as an H-graph using  the lattice coordinates as node features, $\vec{o}_i=(x_i,y_i)$ with $x_i,y_i\in\{-1,0,1\}$ and edge features $\vec{e}_{ij}=(1,1,J^{zz}_{ij})$. EGATE uses 9 EGAT layers, and its decoder is a one hidden layer MLP with 45 hidden units. EGATE is trained with mini-batches (batch size 5) for 50 steps, using an initial learning rate of 0.001 and a step decay of 0.8 after 40 steps.

NNVQE takes $(J^{zz}_1,J^{zz}_2)$ as input, whereas EGATE-NNVQE uses a graph-level embedding $\vec{g}$ with 8-dimensional latent vector. We use a 9-qubit ladder-wise HEA with depth $D=2$, resulting in 117 ansatz parameters. NNVQE is trained with a learning rate of 0.003 for up to 200 iterations, without mini-batch training or a learning rate scheduler.



\paragraph{$H_{\mathrm{XYZ}}^{3\times3}$}: 
We simulate the 9-qubit Hamiltonian $H_{\mathrm{XYZ}}^{3\times 3}$ (Eq.~(\ref{eq:2D_XYZ})), defined on a $3\times 3$ (2D) lattice, using NNVQE with two hidden layers. For both NNVQE and EGATE-NNVQE, the training set for the tunable couplings $J^{yy}$ from 5 equally spaced points in $[-1,1]$ and $J^{zz}_1,J^{zz}_2$ from 9 equally spaced points in $[-2,2]$, yielding 405 training instances. For testing, we use 20 and 40 equally spaced points in $[-4,4]$ and $[-7,7]$ for $J^{yy}$ and $(J^{zz}_1,J^{zz}_2)$, respectively, resulting in 32,000 test instances.

Hamiltonian is encoded as an H-graph using  the lattice coordinates as node features, $\vec{o}_i=(x_i,y_i)$ with $x_i,y_i\in\{-1,0,1\}$ and edge features $\vec{e}_{ij}=(1,1,J^{zz}_{ij})$. EGATE uses 5 EGAT layers, and its decoder is a one hidden layer MLP with 45 hidden units. EGATE is trained with mini-batches (batch size 5) for 50 steps, using an initial learning rate of 0.001 and a step decay of 0.8 after 40 steps (i.e., only the last 10 steps use a different learning rate).

NNVQE takes $(J^{yy},J^{zz}_1,J^{zz}_2)$ as input, whereas EGATE-NNVQE uses a graph-level embedding $\vec{g}$ with 8-dimensional latent vector. We use a 9-qubit ladder-wise HEA with depth $D=2$, resulting in 117 ansatz parameters. EGATE is trained with mini-batches (batch size 20) for 150 steps, using a learning rate of 0.003 without a scheduler.

In the 2D Hamiltonian experiments, we selected $L$ from ${5,7,9}$ based on the configuration that achieved the lowest test loss; however, the differences across these choices were not substantial.



\subsection{Barren Plateau}
To compare barren plateau (BP) behavior between NNVQE and EGATE-NNVQE, we measure how the variance of the ansatz-parameter gradients decays with increasing problem size. We consider the 1D XXZ Hamiltonian $H_{\mathrm{XXZ}}$ with $J^{zz}=1$. We vary the number of qubits from $n=3$ to $9$ and use a ladder-wise HEA whose block depth increases from $D=1$ to $7$, yielding ansatz parameter counts $[24,52,90,138,196,264,342]$, respectively. Standard VQE uses the same ansatz structure, with initial parameters sampled from the uniform distribution, $\mathrm{Unif}[0,2\pi]$. For NNVQE baseline and EGATE-NNVQE, NN weights are initialized from a standard normal distribution $\mathcal{N}(0,1)$. EGATE uses 3 EGAT layers and a one hidden layer decoder MLP with 16 hidden units for all $n(D)$ configurations. For each configuration, EGATE is trained on the fixed Hamiltonian until the reconstruction loss falls below $10^{-5}$.
To isolate the BP trend, we run 500 independent trials for each $n(D)$. In each trial, we compute the gradients at the first optimization step (before any weight updates), i.e., without performing training. Therefore, learning-strategy hyperparameters such as the learning rate are not involved.

\section{Experimental Result}
\label{appendix:sec:generalization}
This section provides the corresponding numerical values in tabular form for the generalization results reported in the main manuscript, Figs.~\ref{fig:1D_generalization_result}  and~\ref{fig:2D_generalization_result}.
For each experiment setting (Hamiltonian family, system size $n$, ansatz depth $D$, and network architecture), we report mean $\pm$ standard deviation over 10 random seeds. For each seed, we compute each metric on the test set, including mean squared error (MSE), mean relative energy error (MRE), and mean fidelity (MF).
All training and evaluating setting is explained in Secs.~\ref{sec:generalization} and~\ref{appendix:sec:experimental setting}.

\begin{table*}[t]
\centering
\renewcommand{\arraystretch}{1.3}

\begin{subtable}{\textwidth}
\centering
\begin{tabular}{|c|c||c|cc|cc|}
\hline
\multicolumn{2}{|c||}{} & \multicolumn{5}{c|}{1 hidden layer} \\
\hline
\multirow{2}{*}{Size} & \multirow{2}{*}{Metric} &
\multicolumn{1}{c|}{NNVQE} &
\multicolumn{2}{c|}{Input-Expanded-NNVQE} &
\multicolumn{2}{c|}{EGATE-NNVQE} \\
\cline{3-7}
& & 
&  & Impr. &  & Impr. \\
\hline

\multirow{3}{*}{\shortstack[c]{$n=4$\\$D=2$\\$(L=5)$}} &
MSE $\downarrow$ &
364.04$\pm$57.32 &
407.32$\pm$59.23 & -11.9\% &
306.15$\pm$101.83 & 15.9\% \\
&
MRE $\downarrow$ &
0.42$\pm$0.04 &
0.46$\pm$0.04 & -9.5\% &
0.36$\pm$0.08 & 14.3\% \\
&
MF $\uparrow$ &
0.55$\pm$0.07 &
0.51$\pm$0.05 & -7.3\% &
0.63$\pm$0.08 & 14.6\% \\
\hline\hline

\multirow{3}{*}{\shortstack[c]{$n=6$\\$D=2$\\$(L=7)$}} &
MSE $\downarrow$ &
802.18$\pm$159.60 &
852.94$\pm$147.64 & -6.3\% &
409.93$\pm$254.54 & 48.9\% \\
&
MRE $\downarrow$ &
0.45$\pm$0.06 &
0.45$\pm$0.06 & 0.0\% &
0.27$\pm$0.11 & 40.0\% \\
&
MF $\uparrow$ &
0.26$\pm$0.02 &
0.27$\pm$0.03 & 3.8\% &
0.36$\pm$0.05 & 38.5\% \\
\hline\hline

\multirow{3}{*}{\shortstack[c]{$n=8$\\$D=2$\\$(L=9)$}} &
MSE $\downarrow$ &
1363.72$\pm$256.52 &
1428.49$\pm$285.05 & -4.7\% &
499.80$\pm$238.92 & 63.4\% \\
&
MRE $\downarrow$ &
0.42$\pm$0.06 &
0.43$\pm$0.07 & -2.4\% &
0.21$\pm$0.06 & 50.0\% \\
&
MF $\uparrow$ &
0.24$\pm$0.03 &
0.24$\pm$0.02 & 0.0\% &
0.35$\pm$0.04 & 45.8\% \\
\hline
\end{tabular}
\caption{1 hidden layer}
\end{subtable}

\begin{subtable}{\textwidth}
\centering
\begin{tabular}{|c|c||c|cc|cc|}
\hline
\multicolumn{2}{|c||}{} & \multicolumn{5}{c|}{2 hidden layer} \\
\hline
\multirow{2}{*}{Size} & \multirow{2}{*}{Metric} &
\multicolumn{1}{c|}{NNVQE} &
\multicolumn{2}{c|}{Input-Expanded-NNVQE} &
\multicolumn{2}{c|}{EGATE-NNVQE} \\
\cline{3-7}

& & &  & Impr. &  & Impr. \\
\hline

\multirow{3}{*}{\shortstack[c]{$n=4$\\$D=2$\\$(L=5)$}} &
MSE $\downarrow$ &
127.37$\pm$70.60 &
158.60$\pm$27.63 & -24.5\% &
74.98$\pm$37.40 & 41.1\% \\
&
MRE $\downarrow$ &
0.20$\pm$0.06 &
0.22$\pm$0.02 & -10.0\% &
0.14$\pm$0.05 & 30.0\% \\
&
MF $\uparrow$ &
0.79$\pm$0.06 &
0.74$\pm$0.08 & -6.3\% &
0.84$\pm$0.06 & 6.3\% \\
\hline\hline

\multirow{3}{*}{\shortstack[c]{$n=6$\\$D=2$\\$(L=7)$}} &
MSE $\downarrow$ &
308.78$\pm$154.64 &
342.06$\pm$146.46 & -10.8\% &
145.16$\pm$97.50 & 53.0\% \\
&
MRE $\downarrow$ &
0.24$\pm$0.07 &
0.23$\pm$0.06 & 4.2\% &
0.15$\pm$0.05 & 37.5\% \\
&
MF $\uparrow$ &
0.37$\pm$0.04 &
0.38$\pm$0.03 & 2.7\% &
0.42$\pm$0.03 & 13.5\% \\
\hline\hline

\multirow{3}{*}{\shortstack[c]{$n=8$\\$D=2$\\$(L=9)$}}&
MSE $\downarrow$ &
620.77$\pm$272.91 &
503.21$\pm$188.61 & 18.9\% &
189.89$\pm$98.55 & 69.4\% \\
&
MRE $\downarrow$ &
0.25$\pm$0.06 &
0.20$\pm$0.05 & 20.0\% &
0.12$\pm$0.03 & 52.0\% \\
&
MF $\uparrow$ &
0.33$\pm$0.04 &
0.37$\pm$0.03 & 12.1\% &
0.41$\pm$0.02 & 24.2\% \\
\hline
\end{tabular}
\caption{2 hidden layers}
\end{subtable}
\caption{Experimental result for generalization experiment on $H_\text{XXZ}$ (Eq.~(\ref{eq:XXZ})), Fig.~\ref{fig:1D_generalization_result}(a). Here, $L$ indicates the number of EGAT layers.  The negative relative improvement (Impr.) indicates worse performance compared to the NNVQE baseline.
}
\label{appendix:table1}
\end{table*}








\begin{table*}[t]
\centering
\renewcommand{\arraystretch}{1.3}
\begin{tabular}{|c|c||cc|c||cc|c|}

\hline
\multicolumn{2}{|c||}{}&
\multicolumn{3}{c||}{1 hidden layer NNVQE}&
\multicolumn{3}{c|}{2 hidden layer NNVQE}\\
\hline

Size&
Metric&
NNVQE&
EGATE-NNVQE&
Impr.&
NNVQE&
EGATE-NNVQE&
Impr.\\ 
\hline

\multirow{3}{*}{\shortstack[c]{$n=4$\\$D=2$\\$(L=5)$}}&
MSE $\downarrow$&
518.08$\pm$99.19&
350.77$\pm$88.51&
32.3\% &
267.89$\pm$96.24&
122.82$\pm$41.40&
54.1\% \\

&
MRE $\downarrow$&
0.52$\pm$0.07&
0.42$\pm$0.06&
19.2\% &
0.35$\pm$0.06&
0.24$\pm$0.04&
31.4\% \\

&
MF $\uparrow$&
0.31$\pm$0.04&
0.33$\pm$0.05&
6.5\% &
0.39$\pm$0.03&
0.42$\pm$0.05&
7.7\% \\\hline\hline

\multirow{3}{*}{\shortstack[c]{$n=6$\\$D=2$\\$(L=7)$}}&
MSE $\downarrow$&
1078.26$\pm$146.87&
528.28$\pm$151.94&
51.0\% &
593.15$\pm$198.27&
291.40$\pm$79.08&
50.9\% \\

&
MRE $\downarrow$&
0.50$\pm$0.03&
0.32$\pm$0.04&
36.0\% &
0.34$\pm$0.07&
0.20$\pm$0.02&
41.2\% \\

&
MF $\uparrow$&
0.22$\pm$0.02&
0.28$\pm$0.03&
27.3\% &
0.31$\pm$0.05&
0.37$\pm$0.02&
19.4\% \\\hline\hline

\multirow{3}{*}{\shortstack[c]{$n=8$\\$D=2$\\$(L=9)$}}&
MSE $\downarrow$&
2113.43$\pm$244.45&
939.59$\pm$246.22&
55.6\% &
892.72$\pm$229.67&
484.31$\pm$131.41&
45.7\% \\

&
MRE $\downarrow$&
0.54$\pm$0.04&
0.33$\pm$0.04&
38.9\% &
0.32$\pm$0.05&
0.20$\pm$0.02&
37.5\% \\

&
MF $\uparrow$&
0.16$\pm$0.02&
0.23$\pm$0.02&
43.8\% &
0.27$\pm$0.04&
0.32$\pm$0.02&
18.5\% \\\hline

\end{tabular}
\caption{Experimental result for generalization experiment on $H_\text{XXZ+X}$ (Eq.~(\ref{eq:XXZ+X})), Fig.~\ref{fig:1D_generalization_result}(b). Here, $L$ indicates the number of EGAT layers.}
\label{appendix:table3}
\end{table*}

\begin{table*}[t]
\centering
\renewcommand{\arraystretch}{1.3}
\begin{tabular}{|c|c||c||c|c|}

\hline
\multicolumn{2}{|c||}{}&
\multicolumn{3}{c|}{2 hidden layer NNVQE}\\
\hline

Size&
Metric&
NNVQE&
EGATE-NNVQE&
Impr.\\ 
\hline

\multirow{3}{*}{\shortstack[c]{$n=9$\\$D=2$\\$(L=9)$}}&
MSE $\downarrow$&
2425.52$\pm$538.46&
1314.44$\pm$334.42&
45.8\%\\

&
MRE $\downarrow$&
0.36$\pm$0.05&
0.30$\pm$0.04&
16.7\%\\

&
MF $\uparrow$&
0.22$\pm$0.02&
0.26$\pm$0.02&
18.2\%\\\hline

\end{tabular}
\caption{Experimental result for generalization experiment on $ H_{\text{XXZ}}^{3\times3}$ (Eq.~(\ref{eq:2D_XXZ})), Fig.~\ref{fig:2D_generalization_result}. Here, $L$ indicates the number of EGAT layers.
}
\label{appendix:table4}
\end{table*}

\begin{table*}[t]
\centering
\renewcommand{\arraystretch}{1.3}
\begin{tabular}{|c|c||c||c|c|}

\hline
\multicolumn{2}{|c||}{}&
\multicolumn{3}{c|}{2 hidden layer NNVQE}\\
\hline

Size&
Metric&
NNVQE&
EGATE-NNVQE&
Impr.\\ 
\hline

\multirow{3}{*}{\shortstack[c]{$n=9$\\$D=2$\\$(L=5)$}}&
MSE $\downarrow$&
1696.87$\pm$301.58&
916.99$\pm$260.71&
46.0\%\\

&
MRE $\downarrow$&
0.46$\pm$0.04&
0.31$\pm$0.04&
32.6\%\\

&
MF $\uparrow$&
0.18$\pm$0.02&
0.23$\pm$0.02&
27.8\%\\\hline

\end{tabular}
\caption{Experimental result for generalization experiment on $ H_{\text{XYZ}}^{3\times3}$ (Eq.~(\ref{eq:2D_XYZ})), Fig.~\ref{fig:2D_generalization_result}. Here, $L$ indicates the number of EGAT layers.}
\label{appendix:table5}
\end{table*}


\end{document}